\documentclass[aps,prb,showpacs,showkeys,twocolumn,superscriptaddress,reprint]{revtex4-1}
\usepackage{natbib}
\usepackage{url}
\usepackage{textcase}
\usepackage{color}
\usepackage{amsmath}
\usepackage[caption=false]{subfig}

\usepackage[normalem]{ulem}
\usepackage[colorlinks=true,citecolor=red]{hyperref}

\newcommand{\unclear}[1]{\textcolor{blue}{\texttt{#1}}}

\newif\ifdraft
\draftfalse 

\newcommand{\GS}[1]{\ifdraft{\color{green}#1}\fi}
\newcommand{\EK}[1]{\ifdraft{\color{magenta}#1}\fi}
\newcommand{\JB}[1]{\ifdraft{\color{cyan}#1}\fi}
\newcommand{\MS}[1]{\ifdraft{\color{red}#1}\fi}

\usepackage{mathtools, bm, wasysym, graphicx, here, upgreek}

\def\D{\mathrm d}
\def\E{\mathrm e}
\def\I{\mathrm i}

\newcommand{\eV}{\,\mathrm{eV}}
\newcommand{\K}{\,\mathrm{K}}
\newcommand{\meV}{\,\mathrm{meV}}
\newcommand{\Ang}{\,\mathrm{\AA}}

\let\vec\bm

\let\epsilon\varepsilon
\let\Pi\varPi
\let\Sigma\varSigma

\def\textsb#1{$_{\text{#1}}$}

\usepackage{xspace}

\makeatletter
\DeclareRobustCommand\SoD{\@ifstar\@SoDcombi\@SoDalone}
\def\@SoDcombi{Star-of-David\xspace} 
\def\@SoDalone{Star of David\xspace} 
\makeatother

\begin{document}

\title{Electronic structure of single layer 1T-NbSe\textsb2: interplay of lattice distortions, non-local exchange, and Mott-Hubbard correlations}

	\author{E. Kamil}
	\affiliation{
		Institut f{\"u}r Theoretische Physik, 
		Universit{\"a}t Bremen, 
		Otto-Hahn-Allee 1, 
		D-28359 Bremen, Germany
	}
	\affiliation{
		Bremen Center for Computational Materials Science, 
		Universit{\"a}t Bremen, 
		Am Fallturm 1a, 
		D-28359 Bremen, Germany
	}
	\author{J. Berges}
	\affiliation{
		Institut f{\"u}r Theoretische Physik, 
		Universit{\"a}t Bremen, 
		Otto-Hahn-Allee 1, 
		D-28359 Bremen, Germany
	}
	\affiliation{
		Bremen Center for Computational Materials Science, 
		Universit{\"a}t Bremen, 
		Am Fallturm 1a, 
		D-28359 Bremen, Germany
	}
	\author{G. Sch\"onhoff}
	\affiliation{
		Institut f{\"u}r Theoretische Physik, 
		Universit{\"a}t Bremen, 
		Otto-Hahn-Allee 1, 
		D-28359 Bremen, Germany
	}
	\affiliation{
		Bremen Center for Computational Materials Science, 
		Universit{\"a}t Bremen, 
		Am Fallturm 1a, 
		D-28359 Bremen, Germany
	}

%
	\author{M. R\"osner}
	\affiliation{
		Department of Physics and Astronomy, 
		University of Southern California, 
		Los Angeles, 
		CA 90089-0484, USA
	}
	\author{M. Sch\"uler}
	\affiliation{
		Institut f{\"u}r Theoretische Physik, 
		Universit{\"a}t Bremen, 
		Otto-Hahn-Allee 1, 
		D-28359 Bremen, Germany
	}
	\affiliation{
		Bremen Center for Computational Materials Science, 
		Universit{\"a}t Bremen, 
		Am Fallturm 1a, 
		D-28359 Bremen, Germany
	}
		\author{G.~Sangiovanni} 
		\affiliation{
			Theoretische Physik 1,
			Julius-Maximilians-Universit\"at W\"urzburg,
			Am Hubland,
			D-97074 W\"urzburg, Germany
		}
	\author{T. O. Wehling}
	\affiliation{
		Institut f{\"u}r Theoretische Physik, 
		Universit{\"a}t Bremen, 
		Otto-Hahn-Allee 1, 
		D-28359 Bremen, Germany
	}
	\affiliation{
		Bremen Center for Computational Materials Science, 
		Universit{\"a}t Bremen, 
		Am Fallturm 1a, 
		D-28359 Bremen, Germany
	}

\date {\today}

\begin{abstract}

Using \textit{ab-initio} calculations we reveal the nature of the insulating phase recently found experimentally in monolayer 1T-NbSe\textsb2. We find soft phonon modes in a large parts of the Brillouin zone indicating the strong-coupling nature of a charge-density-wave instability. Structural relaxation of a  $\sqrt{13}\times\sqrt{13}$  supercell reveals a \SoD* reconstruction with an energy gain of $60\meV$ per primitive unit cell.
The band structure of the distorted phase exhibits a half-filled flat band which is associated with orbitals that are delocalized over several atoms in each \SoD. By including many-body corrections through a combined GW, hybrid-functional, and DMFT  treatment, we find the flat band to split into narrow Hubbard bands. The lowest energy excitation across the gap turns out to be between itinerant Se-$p$ states and the upper Hubbard band, determining the system to be a charge-transfer insulator. Combined hybrid-functional and GW calculations show that long-range interactions shift the Se-$p$ states to lower energies. Thus, a delicate interplay of local and long-range correlations determines the gap nature and its size in this distorted  phase of the monolayer 1T-NbSe$_2$.





\end{abstract}

\maketitle


\section{Introduction}
Two-dimensional (2D) materials often exhibit properties that are distinct from those of their bulk counterparts. The electronic structure and phase diagrams can strongly depend on the thickness of the material, as evident from findings such as the Dirac cone in graphene,\cite{novoselov_two-dimensional_2005} a change from indirect to direct semiconductors,\cite{mak_atomically_2010} increasing and decreasing critical temperatures of superconductivity,\cite{ge_superconductivity_2015,xi_strongly_2015,costanzo_gate-induced_2016} or possibly layer-dependent charge-density waves (CDWs).\cite{calandra_effect_2009,yu_gate-tunable_2015} Due to the all-surface nature of 2D materials, the substrates that they are placed on during growth and in experiments can also have a strong influence on the electronic properties. 

In layered materials, the interplay between strong electron-electron and electron-phonon interactions can lead to rich series of many-body instabilities including Mott-insulating, CDW,  and superconducting phases. 1T-TaS$_2$ presents a classical example in this respect,\cite{Wilson_Salvo_rev1975,Fazekas_TaS2_79,Fazekas_TaS2_80} where a Mott-insulating state emerges within the commensurate $\sqrt{13}\times\sqrt{13}$  CDW in the bulk. Whether or not this CDW and Mott states are realized down to the monolayer thickness is currently a matter of debate.\cite{yu_gate-tunable_2015,albertini_zone-center_2016,Pierre2014}

\begin{figure}
    \includegraphics[width=\linewidth]{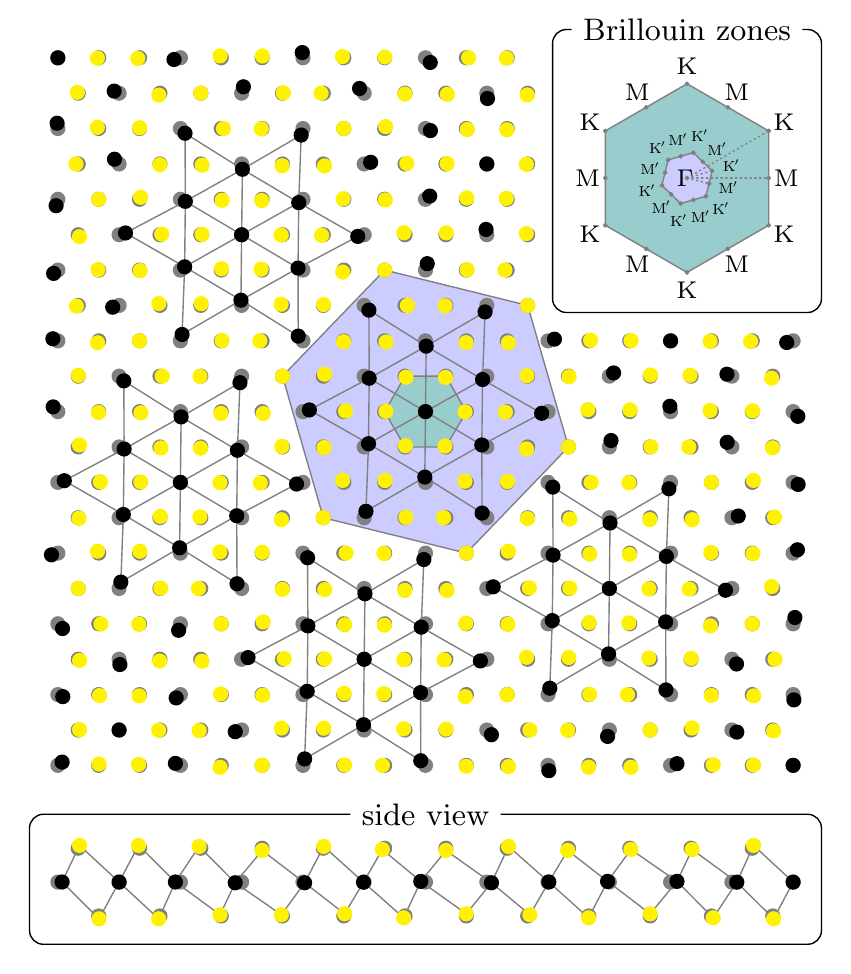}
    \caption{Atomic positions in CDW (black and yellow dots for Nb and Se) and symmetric phase (underlying gray dots). Unit cells and associated Brillouin zones have the same color.}
    \label{fig:cdw_vs_undistorted}
\end{figure}

TaS$_2$ is just one representative of the class of transition-metal dichalcogenides (TMDCs) $MX_2$ with a transition metal $M$ and a chalcogen $X$. The TMDCs occur in two common polymorphs, namely the 1T and 2H structures that consist of octahedral and trigonal-prismatic units, respectively.
An isoelectronic partner of TaS$_2$ is NbSe$_2$. While this material usually crystallizes in the 2H phase, Nakata et al.\cite{Nakata2016} recently succeeded in synthesizing a monolayer of 1T-NbSe$_2$ which was grown epitaxially on bilayer graphene. Their experiments suggested that unlike 2H-NbSe$_2$, which is a metal and hosts a charge-density-wave phase with a $3\times3$ periodicity both in bulk and monolayer form,\cite{ugeda_characterization_2016} 1T-NbSe$_2$ realizes a CDW with$\sqrt{13}\times\sqrt{13}$ periodicity and \SoD--like cluster formation (c.f. Fig.\ref{fig:cdw_vs_undistorted}). In this distorted phase, monolayer 1T-NbSe$_2$ turned out to be an insulator with a gap of around $0.4 \eV$, i.e., this material is phenomenologically similar to 1T-TaS$_2$. 


 
 
In this paper, we provide a first-principles-based account of the electronic structure of monolayer 1T-NbSe$_2$, including electron-electron interaction effects and lattice distortions. We confirm that monolayer 1T-NbSe$_2$ can indeed realize an insulating phase once a \SoD--like CDW is formed.
We show that a delicate interplay of local and long-range electronic interaction effects determines the size and nature of the insulating gap. 

The paper is organized as follows. We start by discussing the electronic structure (Sec.~\ref{sec:undistorted}) and lattice dynamics (Sec.~\ref{sec:distorted}) of the undistorted 1T-NbSe$_2$ phase and investigate the influence of non-local exchange through GW and hybrid-functional calculations. Furthermore, in Sec.~\ref{sec:distorted} we calculate a relaxed structure of a $\sqrt{13}\times\sqrt{13}$ supercell, finding a \SoD* reconstruction. We discuss the electronic structure of the distorted phase in Sec.~\ref{sec:distorted_el}, where we introduce an \textit{ab-initio} estimate of the local Coulomb interaction 
for the flat band and find an opening of a charge-transfer gap in dynamical mean-field theory (DMFT).

 


 \section{Methods}
 We investigate the electronic structure of monolayer 1T-NbSe$_2$ by means of density-functional theory (DFT) with semilocal and hybrid functionals as well as many-body perturbation theory employing the GW approximation. All technical details of these simulations are given in Appendix~\ref{sec:app_DFT}.
The lattice dynamics is studied using density-functional perturbation theory (DFPT) as detailed in Appendix~\ref{sec:app_DFPT}. The \textit{ab-initio} estimation of electron-electron interaction is provided in Appendix~\ref{sec:coloumbestimation}. Finally, the way we combine first-principles calculations with dynamical mean-field theory (DMFT) to study the effects of local correlations is discussed in Appendix~\ref{sec:DMFT}.
 


\begin{figure*}
\begin{minipage}{.5\linewidth}
\centering
\subfloat[]{\label{fig:undis_band_nbse2}%
\includegraphics[width=1\linewidth]{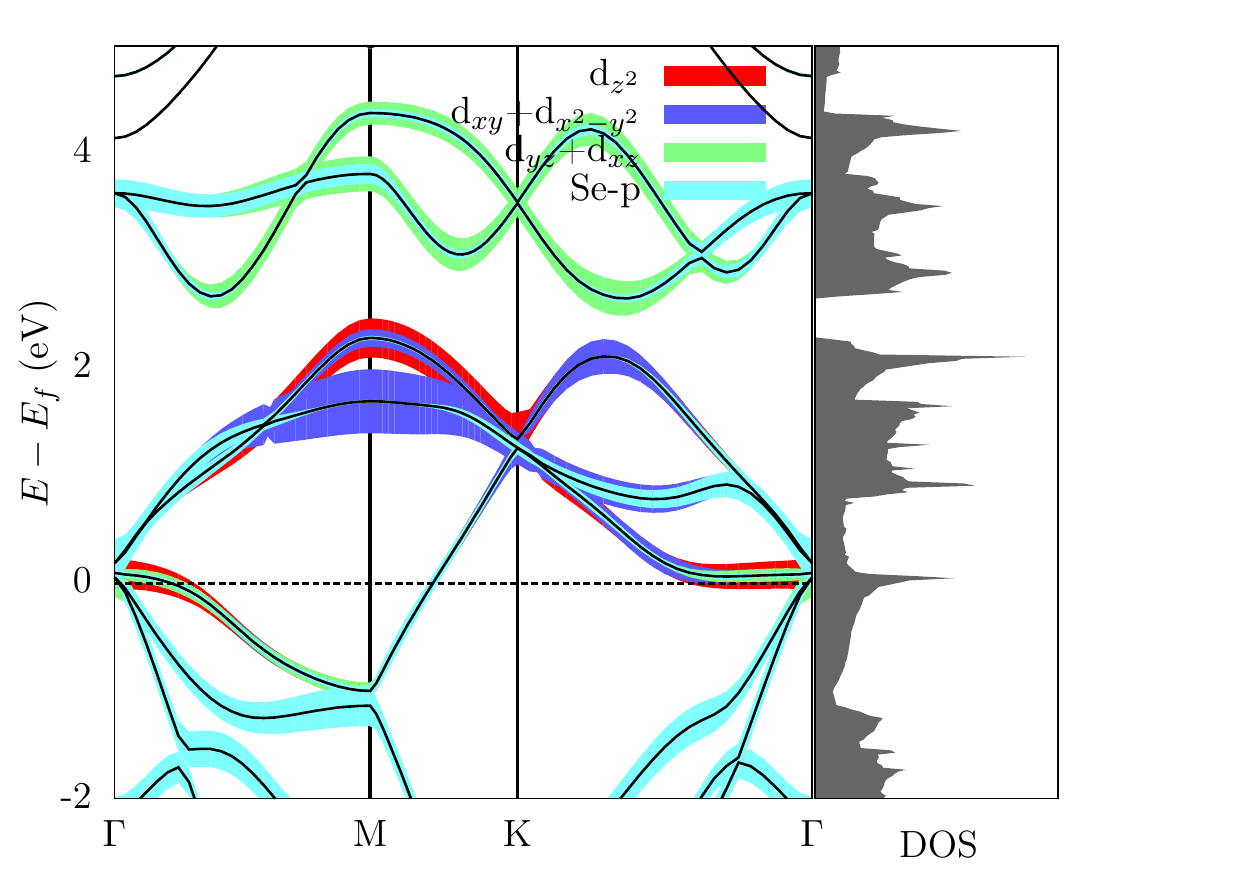}}%
\end{minipage}\hspace{0cm}%
\begin{minipage}{.5\linewidth}
\centering
\subfloat[]{\label{fig:undis_band_nbse2_gw_hse06}%
\includegraphics[width=1\linewidth]{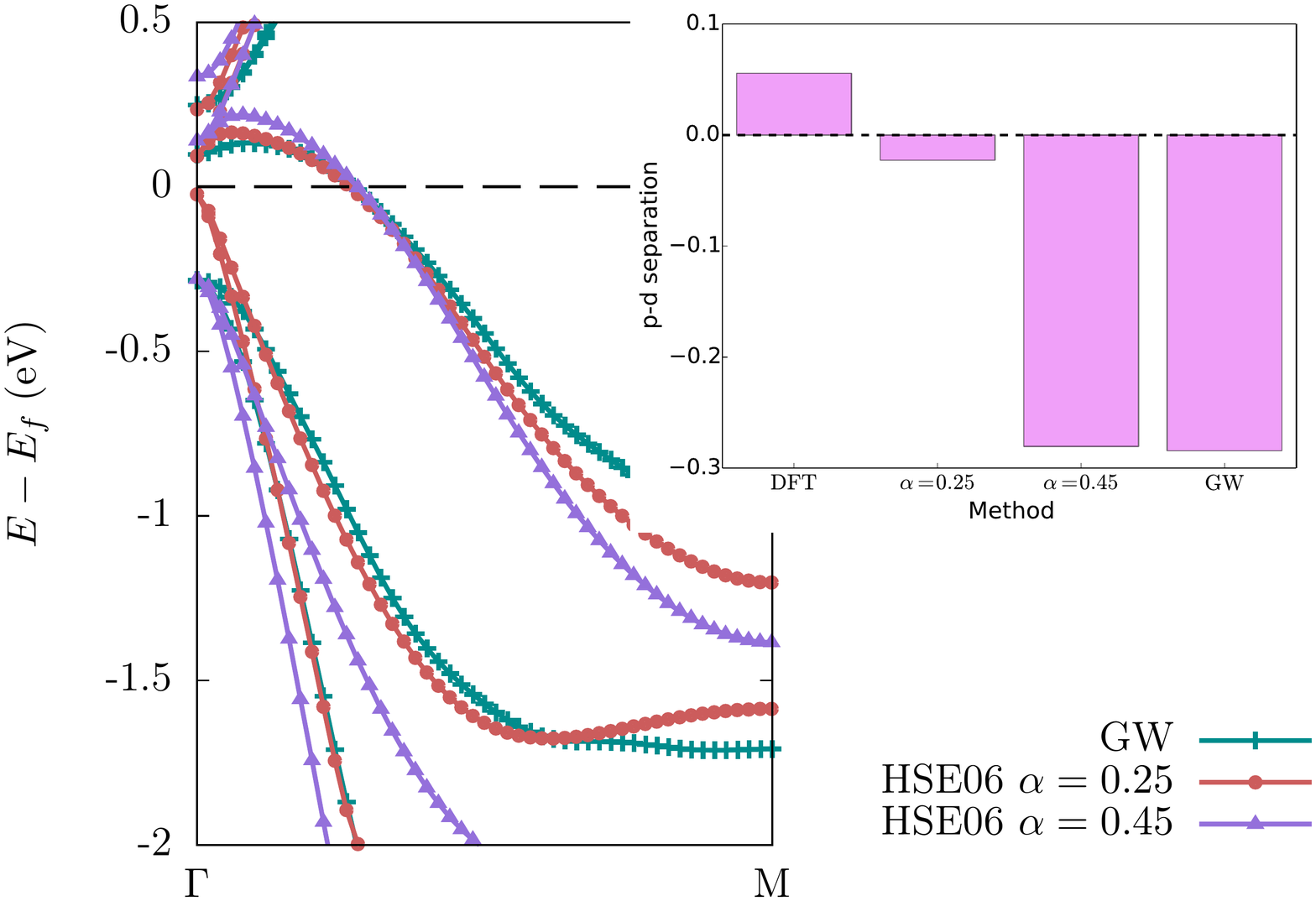}}%
\end{minipage}
\caption{(a) Non-spin-polarized band energies and orbital-projected density of states (color-shaded region) along the path $\Gamma$-M-K-$\Gamma$ containing high-symmetry points in the Brillouin zone corresponding to the primitive unit cell of the undistorted monolayer 1T-NbSe$_2$. The right part corresponds to the total density of states. (b) Non-spin-polarized GW and HSE06 band energies along the path $\Gamma$-M. The bar graph shows the position of Se-$p$-like bands at the $\Gamma$ point relative to the Fermi energy, obtained from GGA, GW and HSE06 functionals.
}
\end{figure*}

\section{Electronic structure of the undistorted phase}
\label{sec:undistorted}
We start by discussing the structural and electronic properties of monolayer 1T-NbSe$_2$ in the undistorted phase. To this end, we consider a simple hexagonal unit cell containing one niobium and two selenium atoms. The structural relaxation using a PBE functional leads to a lattice constant of $a=3.49 \Ang$ and a vertical distance of $z=\pm 1.67 \Ang$ of the selenium atoms from the transition-metal (Nb) plane. Next, the electronic properties were calculated for the relaxed lattice parameters. In Fig.~\ref{fig:undis_band_nbse2}, the band energies and the orbital characters are plotted along the Brillouin-zone path $\Gamma$-M-K-$\Gamma$ connecting the high-symmetry points. As evident, DFT with PBE density functional, predicts 1T-NbSe$_2$ to be a metal with high density of states at the Fermi level. The bands in the range from $0$ to  $2 \eV$ above the Fermi level have mostly Nb-$d_{z^2}$, -$d_{xy}$ and -$d_{x^2-y^2}$ character. Bands in the range from $-1$ to $0 \eV $ below the Fermi level have Se-$p$ orbital character. The Se-$p$ bands are not well separated  and hybridize strongly with Nb-$d$ bands around the $\Gamma$ point near the Fermi level, which distingiushes the electronic structure of 1T-NbSe$_2$ from 1T-TaS$_2$, where the Se-$p$ bands are well separated from the Fermi level.\cite{Pierre2014}  The high DOS at the Fermi level in the monolayer 1T-NbSe$_2$ is an indication of possible electronic instabilities. 

\begin{figure}
    \includegraphics[width=\linewidth]{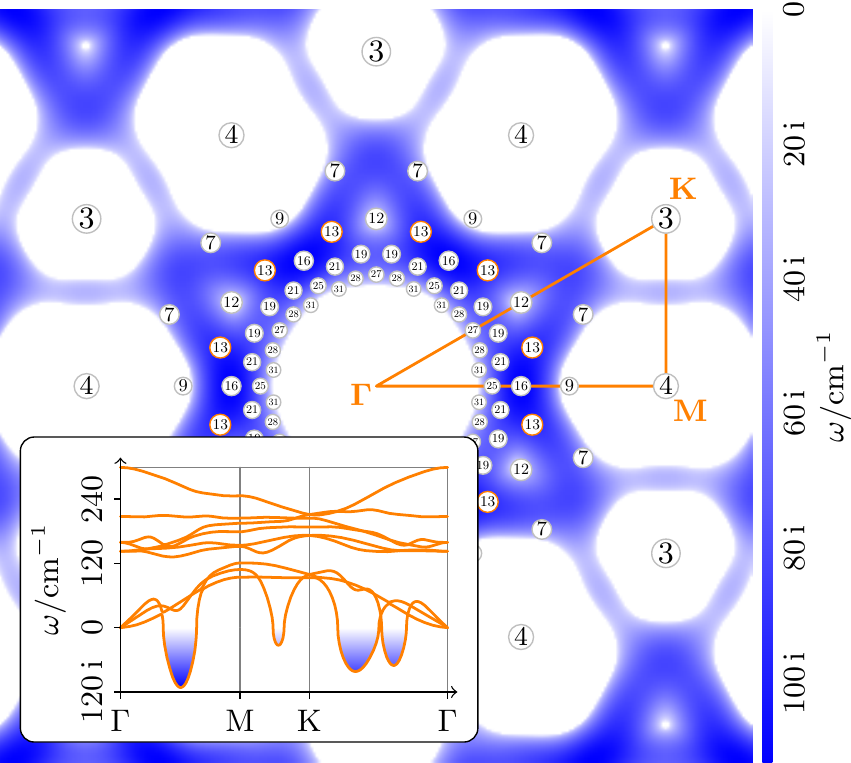}
    \caption{Distribution of unstable phonon modes in reciprocal space, as obtained from DFPT for undistorted 1T-NbSe$_2$. The magnitude of imaginary frequencies is displayed in shades of blue. The wave vectors belonging to potential commensurate $\sqrt N \times \sqrt N$  CDWs are marked with encircled numbers $N$. Inset: Full dispersion along high-symmetry path.}
    \label{fig:phonon_dispersion}
\end{figure}

\subsection{Effects of non-local exchange on Se-\textit p bands.}

The hybridization of the downward-dispersing Se-$p$ bands with the Nb-$d$ bands at the $\Gamma$ point is strongly affected by non-local exchange effects.  In Fig.~\ref{fig:undis_band_nbse2_gw_hse06} the band structures obtained from GW and HSE06 approximations for the undistorted structure are plotted along the path $\Gamma$-M. Unlike the PBE functional which leads to Se-$p$ bands starting $\sim 55 \meV$  above the Fermi level around the $\Gamma$ point (see Fig.~\ref{fig:undis_band_nbse2}), the non-local exchange accounted for in the GW approximation shifts the Se-$p$ bands $\sim 0.28\eV$ below the Fermi level such that they are separated from the Nb-$d$ bands. The GW prediction of the position of downward-dispersing Se-$p$  bands around the $\Gamma$ point is in agreement with the ARPES results for the \SoD* phase from Ref.~\onlinecite{Nakata2016}. 

In order to be able to study the effect of (screened) non-local exchange also in the distorted phase, where GW calculations are numerically too demanding, we turn to hybrid-functional calculations and compare band structures of the undistorted phase obtained with different flavors of HSE06 to those from GW (Fig.~\ref{fig:undis_band_nbse2_gw_hse06}). The traditional admixture of $\alpha=0.25$ exact Fock exchange reproduces the shape of the GW bands in most parts of the $\Gamma$-M path quite well, except for the maximimum of the Se-$p$ bands right at $\Gamma$, where the top of the Se-$p$ bands is already below the Fermi level by $\sim 22\meV$. To obtain the same separation within the HSE06 approximation, a large admixing $\alpha=0.45$ of the exact Fock exchange is required.  However, in the case of  $\alpha=0.45$ the Se-$p$ bands are much further below the Nb-$d$ bands in most parts of the $\Gamma$-M path. Hence, the match between HSE06 and GW is best for the traditional value of $\alpha=0.25$, which we will also employ in the following.


\begin{figure*}
\begin{minipage}{.5\linewidth}
\centering
\subfloat[]{\label{fig:dis_band_nbse2}%
\includegraphics[width=1\linewidth]{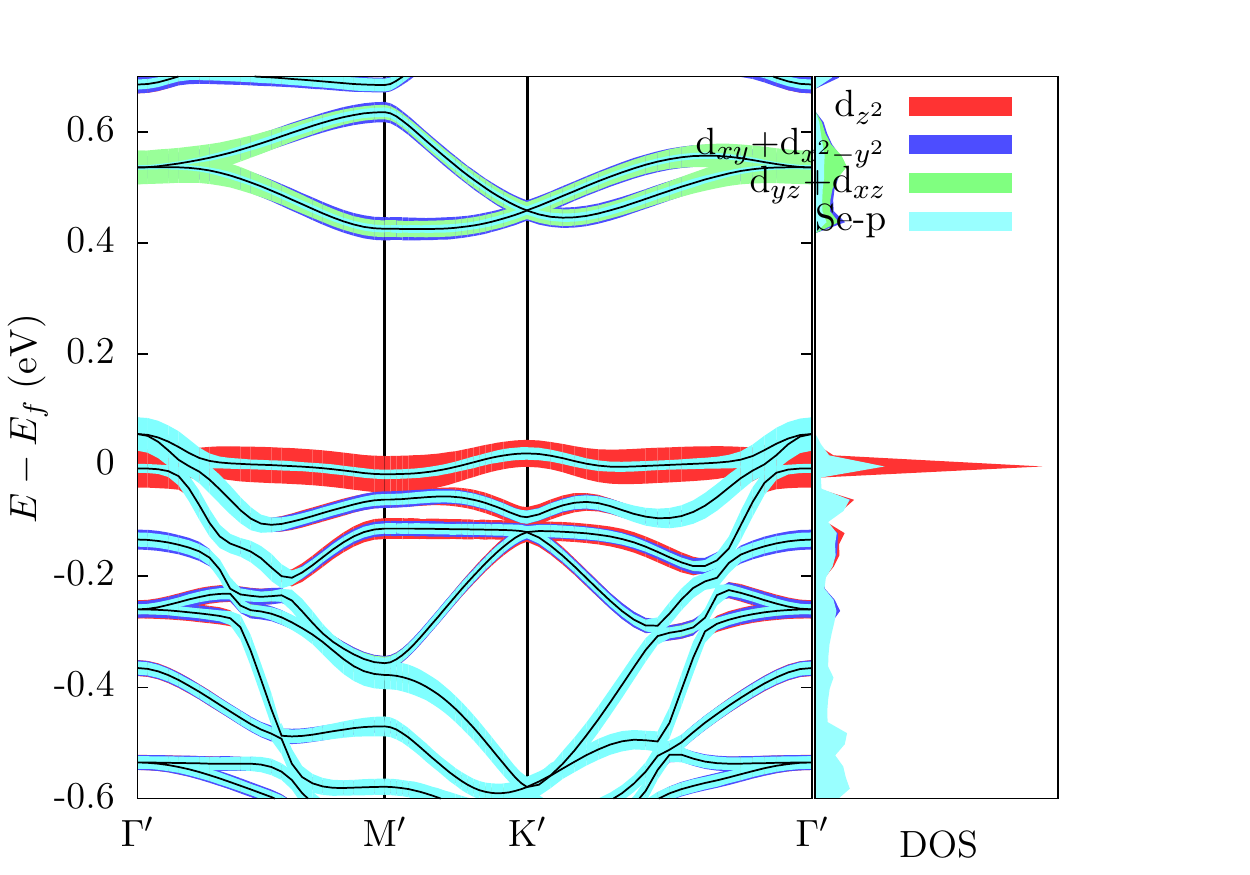}}%
\end{minipage}\hspace{0.cm}%
\begin{minipage}{.5\linewidth}
\centering
\subfloat[]{\label{fig:dis_band_hybrid}%
\includegraphics[width=1\linewidth]{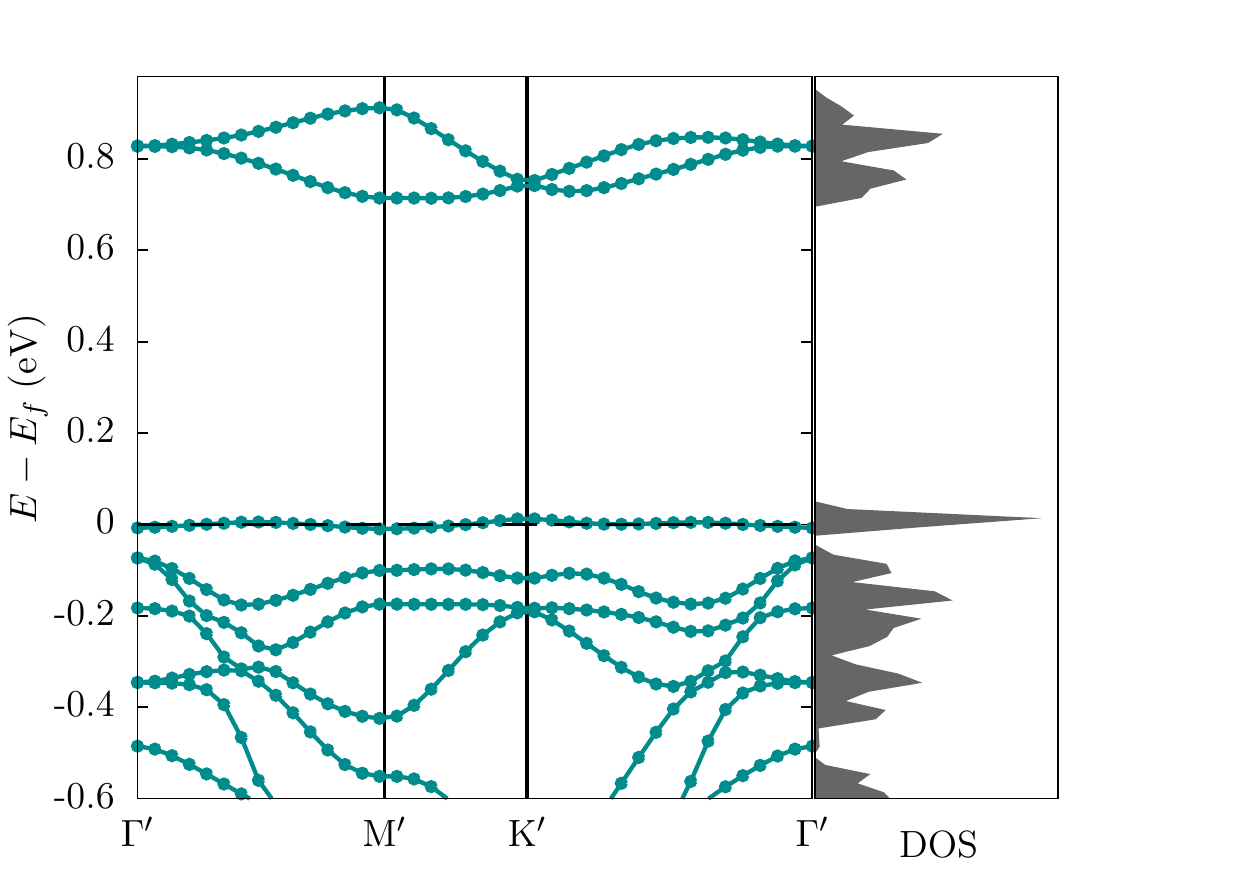}}%
\end{minipage}
\begin{minipage}{0.5\linewidth}
\centering
\subfloat[]{\label{fig:unfolded_band}%
\includegraphics[width=1\linewidth]{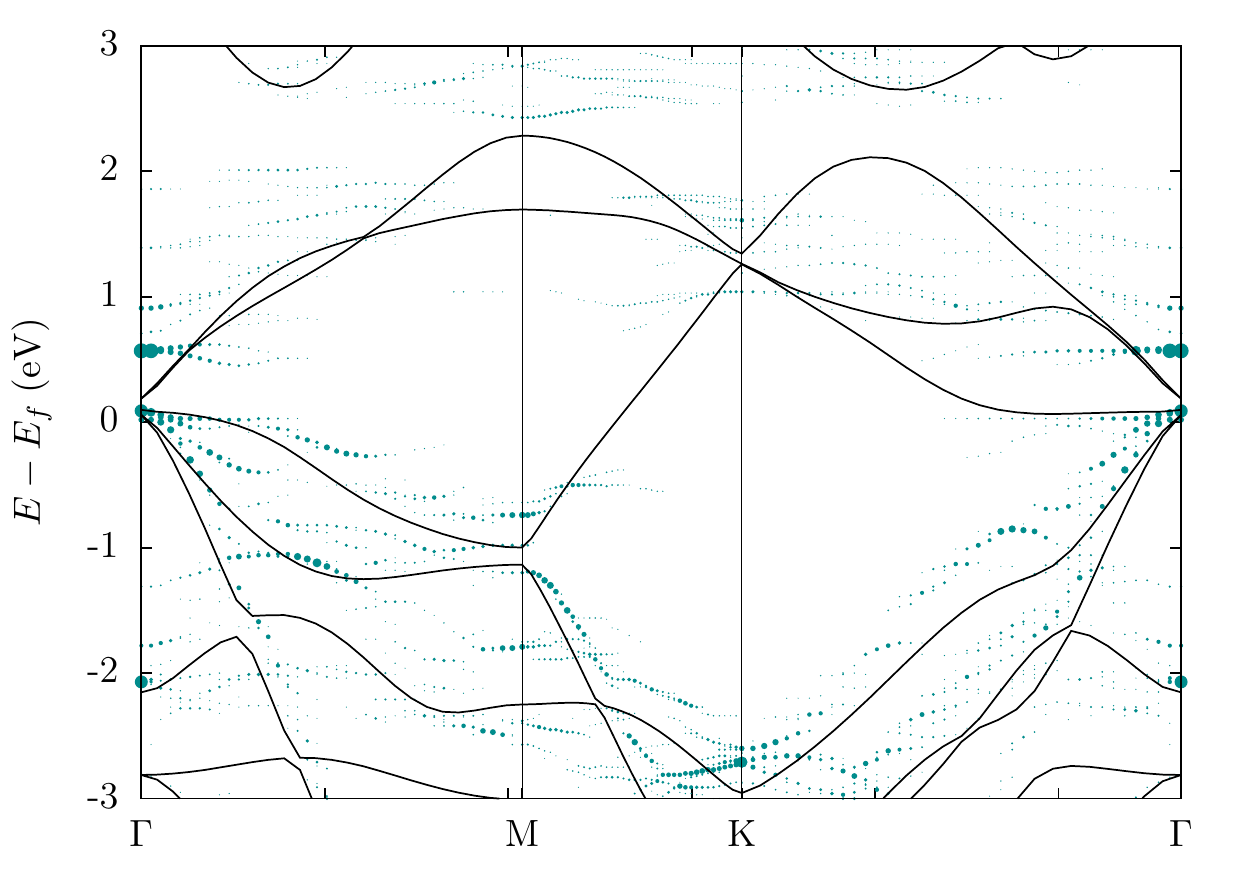}}%
\end{minipage}%
\begin{minipage}{0.5\linewidth}
\centering
\subfloat[]{\label{fig:parchg}%
\includegraphics[width=1.0\linewidth]{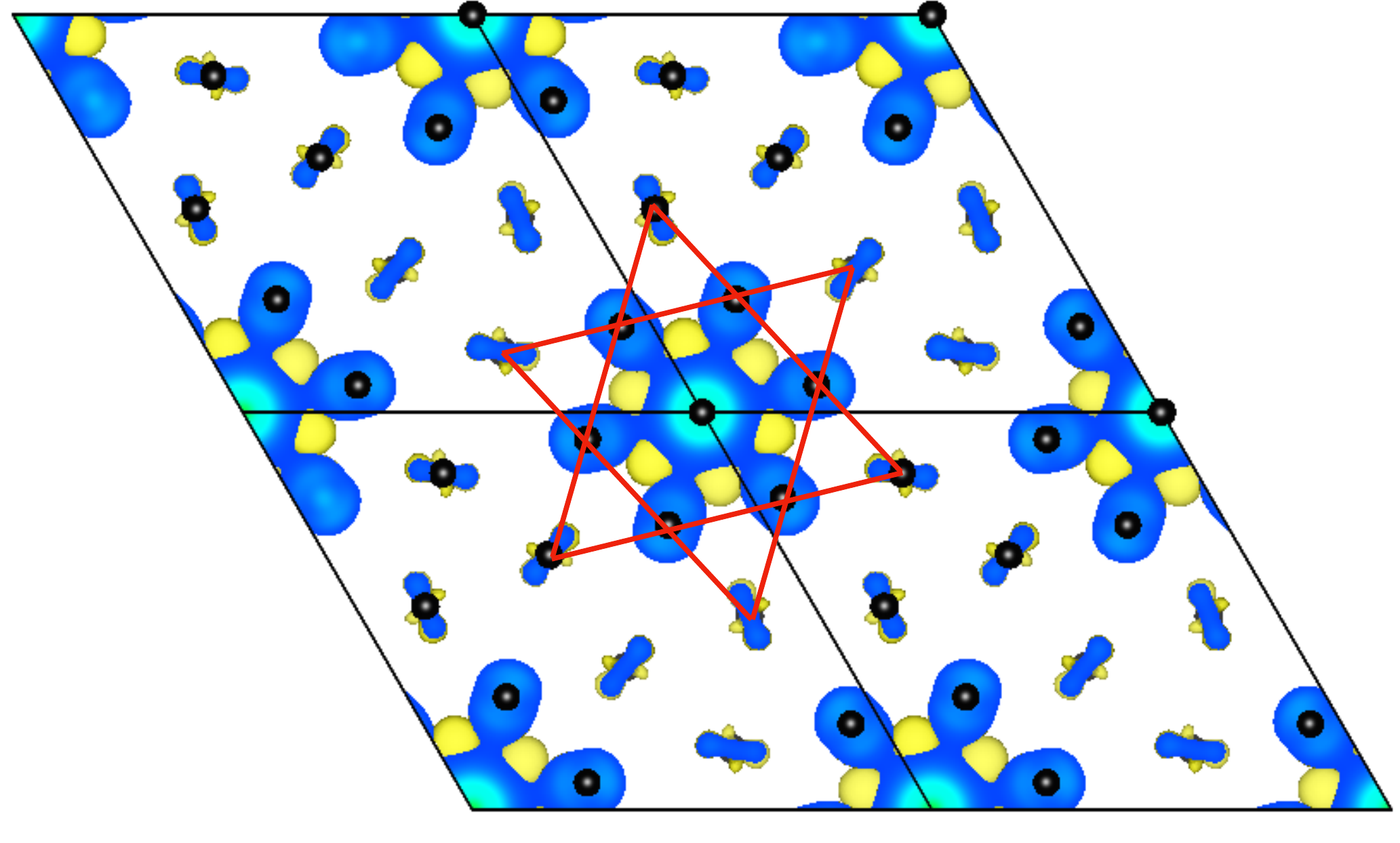}}
\end{minipage}
\caption{The band structure for monolayer 1T-NbSe$_2$ in the distorted CDW phase obtained from non-spin-polarized (a) GGA and (b) HSE06 ($\alpha=0.25$) calculations, along the path $\Gamma$-M'-K'-$\Gamma$ connecting the high-symmetry points of the Brillouin zone corresponding to the \SoD* supercell. (c) Solid dots: The band structure in the CDW phase unfolded to the primitive Brillouin zone. The size of the dots corresponds to the spectral weight. Solid lines: The band structure of the undistorted lattice. (d) Partial charge density associated with the flat band at the Fermi level. The solid black circles correspond to the position of niobium atoms. The predominant part of the charge density is concentrated on the atoms in the center and the inner ring of the \SoD.}
\label{fig:distorted}
\end{figure*}

\section{Lattice distortion}
\label{sec:distorted}

The 1T-NbSe$_2$ islands in the experiment were reported\cite{Nakata2016} to host a CDW with $\sqrt{13}\times\sqrt{13}$ periodicity and a corresponding formation of \SoD* clusters as shown in Fig~\ref{fig:cdw_vs_undistorted}.

To detect possible lattice instabilities from first principles, we calculate the phonon dispersion of the undistorted structure via DFPT, which is displayed in the inset of Fig.~\ref{fig:phonon_dispersion}. It features imaginary frequencies of in-plane acoustic modes in large parts of the reciprocal space, better seen from the color plot in the main panel of the figure, which indicate structural instability. The wave vectors belonging to a $\sqrt 13 \times \sqrt 13$ CDW are located deep inside this instable region. By contrast, there is no indication of the $3 \times 3$ CDW found in many 2H polymorphs.\cite{Uge15}

To accout for the \SoD--like cluster formation, we consider a supercell (see Fig.~\ref{fig:cdw_vs_undistorted}) with $\sqrt{13}\times\sqrt{13}$ periodicity and perform structural relaxation within the GGA approximation for the density functional.
As evident from Fig.~\ref{fig:cdw_vs_undistorted}, with respect to the undistorted lattice, the optimized structure reveals an inward displacement of niobium atoms within the transition-metal plane towards the central atom of the \SoD* configuration. The energy gain through a \SoD--like reconstruction is $\sim 60 \meV$ per primitive unit cell. 

To summarize, the soft phonon modes in a large part of the Brillouin zone indicate towards a strong-coupling nature of the charge-density wave instability, as opposed to a Fermi-surface-nesting scenario, and the emergence of commensurate CDW with \SoD* reconstruction is associated with a sizeable energy gain when compared with the undistorted phase.


\section{Electronic properties of the distorted phase}
\label{sec:distorted_el}


The theoretical description of the insulating behavior of the distorted phase of the monolayer 1T-NbSe$_2$ requires the inclusion of electronic correlations beyond DFT. Therefore, we now study the electronic properties of the \SoD--distorted phase at different levels of approximation.

In Fig.~\ref{fig:dis_band_nbse2}, the GGA band structure of the distorted lattice is plotted along the path $\Gamma$-M$'$-K$'$-$\Gamma$ connecting the high-symmetry points in the Brillouin zone corresponding to the $\sqrt{13}\times\sqrt{13}$ supercell. At the Fermi level, there is a flat band with predominantly Nb-$d_{z^2}$ character which hybridizes with Se-$p$ bands around the $\Gamma$ point. The partial charge density corresponding to the flat band is plotted in Fig.~\ref{fig:parchg} from which it is evident that the flat band is associated with orbitals that are delocalized over several atoms in each \SoD. The maximum contribution arises from niobium atoms in the center and the inner ring of the \SoD. This observation will be used in the following subsection to obtain an \textit{ab-initio} estimate of the effective Hubbard type interaction for the flat band.

Next, we study the effects of non-local exchange on the hybridization of the flat band with Se-$p$-like bands around the $\Gamma$ point. As evident from Fig~\ref{fig:dis_band_hybrid}, the hybridization is strongly suppressed with the inclusion of non-local exchange effects through the HSE06 functional ($\alpha=0.25$). The position of the band lying below the flat band is rather sensitive to the exact-exchange admixing parameter $\alpha$. In the following subsection, when presenting DMFT calculations for the relevant Nb-$d$ and Se-$p$ orbitals, we will show that the position of the $p$ bands indeed affects the size and nature of the insulating gap. 

In Fig.~\ref{fig:unfolded_band}, we have unfolded \cite{bandup1,bandup2} the distorted GGA band structure to the primitive Brillouin zone. There is a considerable reduction of spectral weight along the edge $\mathrm{M-K}$ and deep inside the Brillouin zone. In addition, we observe the opening of a CDW gap $\sim ~0.4\eV$ above the Fermi level. 

So far we have seen that, though acting in the right direction to explain experiments, neither the CDW order nor the non-local exchange makes the system insulating. 
We therefore improve our treatment of local many-body effects by means of DMFT. The goal of the next subsection is hence to see if this yields a correct description of the insulating nature of monolayer 1T-NbSe\textsb2 in the distorted phase.

%


\subsection{Possibility of Mott Physics}

The following GGA+DMFT calculations allow us to quantify the relative importance of two competing effects involving the $d_{z^2}$-derived flat band close to $E_\text{F}$. The first is the electron-electron correlation on the $d_{z^2}$-orbital. The second is the single-particle hybridization between this orbital and the rest of the system, in particular Se-$p$ ones.
On the one hand, it is well known that ``$d$-only'' DMFT calculations have a strong tendency to yield Mott-insulating phases close to integer fillings. On the other hand, when ``ligand'' orbitals are explicitly included in the basis set used for DMFT,  this tendency is drastically reduced, due to important charge fluctuations between the $d$ and the $p$ manifolds. This is mitigated by introducing double-counting counter-terms and non-local interaction terms. We will investigate both effects and determine whether or not in the case of 1T-NbSe$_2$ the tendency towards a Mott-insulating solution remains strong, even in the presence of the $d$-$p$ hybridization derived from first principle. Computational details are described in Appendix.~\ref{sec:DMFT}.

We start with the estimation of the Hubbard-type interaction for the flat band from first principles.
As mentioned in the previous section, the character or the flat band is associated with orbitals delocalized over several atoms in each \SoD. 
Therefore, to obtain an estimate for the Hubbard-type interaction, we evaluate the screened Coulomb interaction for the undistorted phase and perform an average over the \SoD. The fact that atoms on the outer ring of the \SoD have a minor contribution to the partial charge density of the flat band (see Fig.~\ref{fig:parchg}) motivates us to perform the  average once excluding and once including the outer ring. The scheme is described in detail in Appendix.~\ref{sec:coloumbestimation} and leads to $U\sim0.39\eV$ ($0.30\eV$) by excluding (including) the outer ring of the \SoD in the averaging procedure. This estimate provides a lower bound for the Hubbard-type interaction, due to the fact that we use the non-constrained random-phase approximation for the screened Coulomb interaction which also includes screening effects from the bands within the correlated subspace. In fact, we would need to perform constrained random-phase-approximation (cRPA) calculations to exclude any screening from the correlated subspace which would increase the effective U. To account for this effect, we use $U=0.4\eV$ in the following.
%

\begin{figure}[h]
    \includegraphics[width=1\linewidth,height=!]{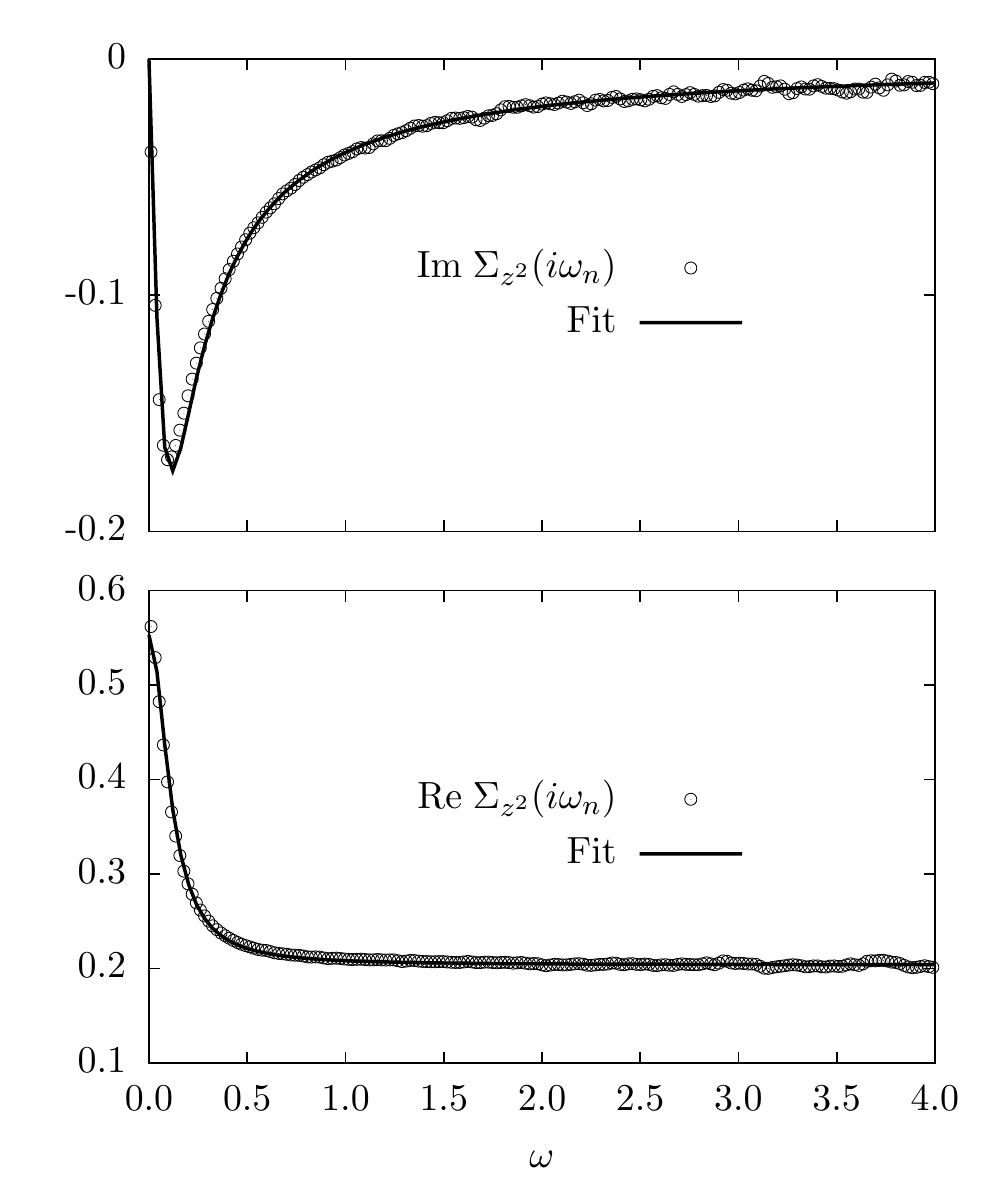}
    \caption{\label{fig:siw_u4}The DMFT data for the imaginary and real part  of the self-energy $\Sigma_{z^2}(\I \omega)$ for the correlated $d_{z^2}$-like orbital for $U=0.4\eV$, $T=40\K$ and the double-counting correction $\mu_{\mathrm{dc}}$ is fixed to $\mathrm{FLL}$. The solid dots correspond to the CT-QMC data whereas the solid line is the corresponding fit function given in Eq.~\eqref{eq:self_energy}. The fit parameters are $\bar{\sigma}=0.203$ and $\omega_0=0.458$. 
    }  
\end{figure}
 
\begin{figure}
     \includegraphics[width=\linewidth]{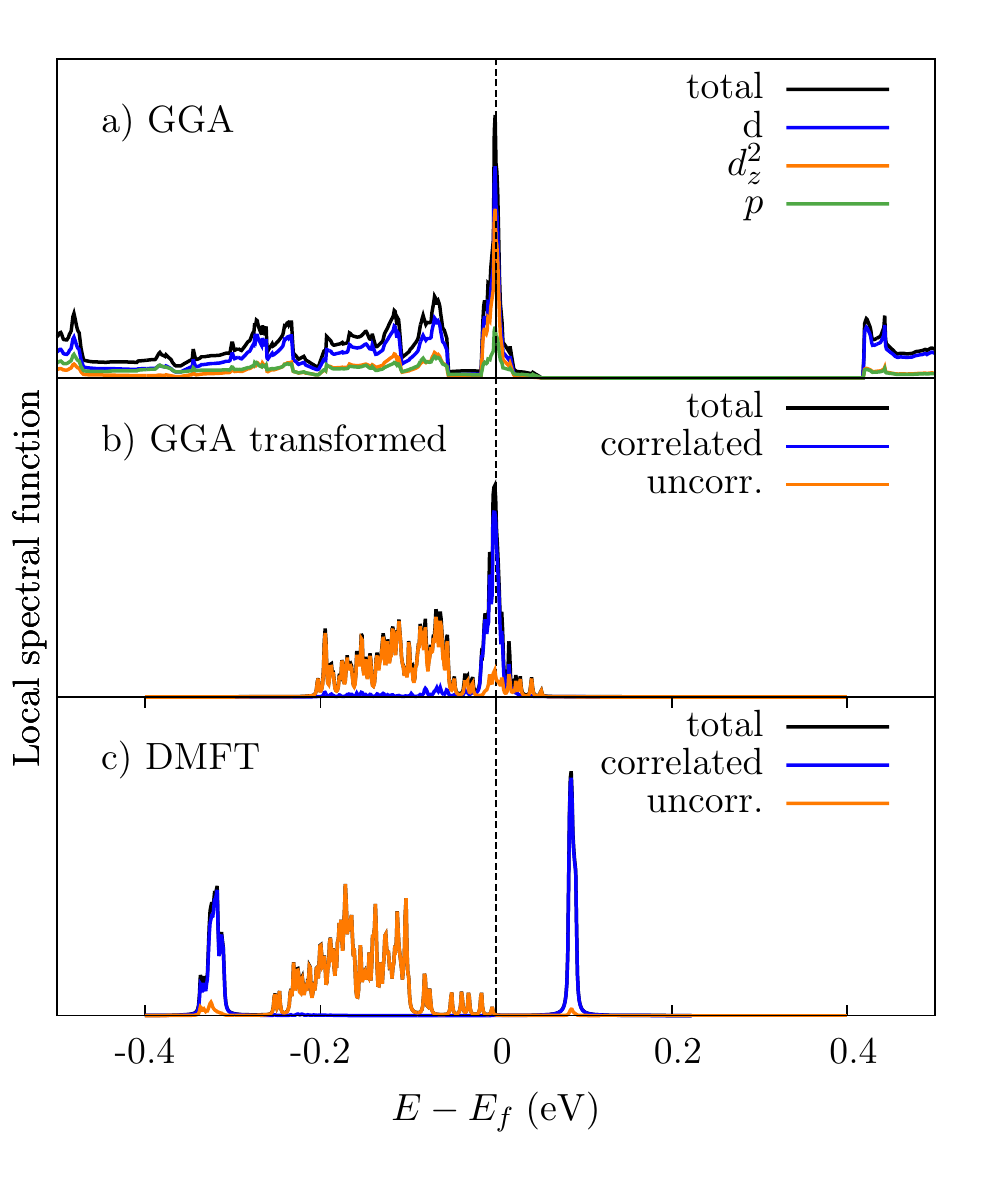}
     \caption{Local spectral function in the distorted (CDW) phase from (a) GGA with partial DOS of $d$ orbitals, only $d_{z^2}$ orbitals, and $p$ orbitals, (b) DOS after unitary transformation with partial DOS of orbitals considered correlated and uncorrelated, and (c) DOS from DMFT calculations with $U=0.4\eV$ at $T=40\K$ and the double-counting correction term $\mu_{\mathrm{dc}}$ is fixed to $\mathrm{FLL}$. }
     \label{fig:dos_DMFT}
 \end{figure}

In the next step we obtain a GGA-based tight-binding model where we include the three bands closest to the Fermi energy. We use a projection onto the $d_{z^2}$ orbitals to define a correlated subspace which is augmented with the Hubbard-type interaction of $U=0.4\eV$. 
We discuss the outcome of this projection in terms of a comparison of the original GGA partial DOS and the DOS of the projection, both shown in Fig.~\ref{fig:dos_DMFT}a-b. In GGA, the states at the Fermi energy are dominantly of $d_{z^2}$ character and the states ranging from $-0.2\eV$ to $-0.05\eV$ are a mixture of $d$ and $p$ states. This is reflected in the projected Hamiltonian, whose spectral weight is entirely distributed among three local orbitals only, the Nb-$d_{z^2}$ and the two ``ligand''-ones, made of a mixture of the other Nb-$d$ and Se-$p$. This set of three orbitals constitutes therefore the basis for our GGA+DMFT calculations. The estimated value of $U=0.4\eV$ is applied in the following to the $d_{z^2}$ orbital only. The other two orbitals are treated as uncorrelated but, being included in the low-energy Hamiltonian, they can exchange charge with the $d_{z^2}$-like one via hybridization processes.
 
Treating orbitals with different degree of correlation in DFT+DMFT requires the inclusion of a double-counting correction, which compensates the DFT contribution. We denote it by $\mu_\text{dc}$ and use the fully-localized limit (FLL).\cite{czyzyk_local-density_1994} As discussed in Sec.~\ref{sec:undistorted}, long-range interactions shift the $p$ bands to lower energies. Since we do not consider any long-range interactions in DMFT, we mimic this behavior by adjusting the position of the uncorrelated bands with respect to the correlated states by an additional energy shift $\Delta_p$ in the projected Hamiltonian.

We solve the DMFT equations self-consistently with a continuous-time quantum Monte Carlo impurity solver using the \emph{w2dynamics} package. \cite{parragh_conserved_2012, wallerberger_w2dynamics:_2018}

In Fig. \ref{fig:siw_u4} we show the self-energy on Matsubara frequencies for the $d_{z^2}$ orbital, with a ``pure'' FLL double counting ($\Delta_p=0$). We fit the numerical data with an analytic expression for a (non particle-hole symmetric) atomic self-energy,
\begin{equation}
\label{eq:self_energy}
 \Sigma^{\rm{fit}}_{z^2}(\I \omega_n)=\bar{\sigma}+\frac{U^2}{4 \I \omega_n+\omega_0}.
\end{equation}
The excellent agreement shows that the system is deep in the atomic limit.

  \begin{figure}
     \includegraphics[width=\linewidth]{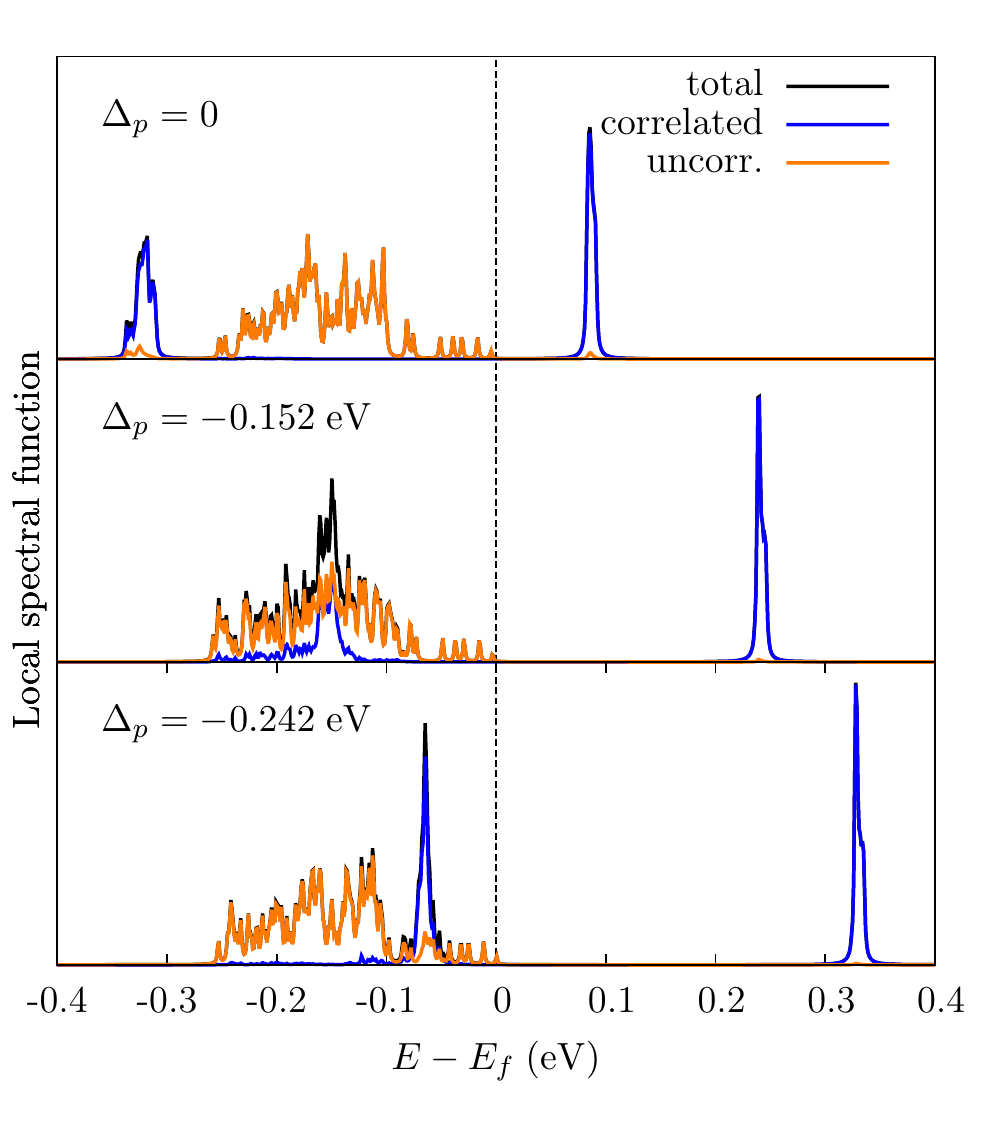}
     \caption{Local spectral function in the distorted (CDW) phase from DMFT calculations with $U=0.4\eV$ at $T=40\K$ for different values of energy shifts $\Delta_p$ for the uncorrelated bands. The double counting correction term $\mu_{\mathrm{dc}}$ for the correlated band is fixed to $\mathrm{FLL}$.
}
     \label{fig:dos_DMFT_shift}
 \end{figure}

The analytic form of the self-energy makes it easily possible to calculate the DMFT spectral function on real frequencies, which we show in Fig.~\ref{fig:dos_DMFT}c. The correlated $d_{z^2}$-like states are split into a sharp upper and lower Hubbard band, separated approximately by the Coulomb interaction $U$. We find the system to be insulating with a gap between $p$-like and $d_{z^2}$-like states of $\sim 0.1\eV$, i.e., we find a charge-transfer insulator.

To mimic the influence of the non-local exchange effects on the $p$-$d$ splitting (c.f. Figs.~\ref{fig:undis_band_nbse2_gw_hse06} and \ref{fig:dis_band_hybrid}) in the DMFT simulations we shift the uncorrelated bands by $\Delta_{p}=-0.15 \eV$ and $\Delta_{p}=-0.24 \eV$ which approximately corresponds to shifts observed in HSE06 ($\alpha=0.25$) and GW/HSE06 ($\alpha=0.45$) calculations respectively. From the resulting spectral functions presented in Fig.~\ref{fig:dos_DMFT_shift}, we observe that the gap grows with $\Delta_p$. The case of $\Delta_{p}=-0.24 \eV$, which closely mimics the $p$-$d$ shift from DFT to GW calculations found in Sec.~\ref{sec:undistorted}, together with a local Coulomb interaction of $U=0.4\eV$, leads to a gap close to the experimentally-observed one of $0.4\eV$ in Ref.~\onlinecite{Nakata2016}. Slightly larger shifts $\Delta_p$ will push the system from a charge-transfer to a Mott insulator.

In our DFT+DMFT approach we have used a low-energy model for DMFT derived after including the contribution of non-local exchange and the CDW symmetry breaking. This choice of the low energy model is a generalization of previous (successful) low-energy treatments of distorted 1T-TaS$_2$\cite{Perfetti_TaS2_2006,Freericks_pruschke_tas_2009} but it is strictly speaking valid only under the assumption that a full DFT+DMFT treatment with the appropriate supercell, non-local corrections and extended basis set would give similar results. Such a prohibitively large calculation goes beyond the purpose of the present work.

\section{Conclusion}
We have investigated monolayer 1T-NbSe$_2$ by combined  DFT, GW, DFPT, and DMFT simulations. With DFT (GGA) simulations, we have shown that there are unstable phonon modes in large parts of the Brillouin zone including those wave vectors corresponding to a $\sqrt{13}\times\sqrt{13}$  commensurate CDW and that formation of this particular CDW with \SoD--like distortion of the lattice is energetically stable with respect to the undistorted phase. 
A very flat band at the Fermi level in the CDW phase renders the system susceptible to a Mott instability in the presence of local Coulomb interaction. A first-principles estimate of the Hubbard-type interaction for the flat band is provided. Subsequent GGA+DMFT $d$-$p$ calculations in the distorted phase yield a correlated insulator deep in the atomic limit. The size and nature of the insulating gap is determined by an interplay of local and long-range Coulomb interactions. For strongly screened non-local interactions, as assumed in GGA we arrive at a charge-transfer insulator, while inclusion of non-local interactions on a GW level rather suggests free-standing 1T-NbSe$_2$ to be at the verge between a Mott and a charge-transfer insulator. Importantly, substrates can be used to effectively tune non-local interactions in 2D materials and to possibly switch 1T-NbSe$_2$ between the clear charge-transfer insulator and a more Mott-Hubbard-type insulator.






\section{Acknowledgement}
Financial support from the Deutsche Forschungsgemeinschaft through RTG 2247 and the computing time from HLRN (project \texttt{hbp00045}) are gratefully acknowledged. We would like to thank Merzuk Kaltak for useful discussions. M.R. would like to thank the Alexander von Humboldt Foundation for support. \\

\textbf{Note}: Two similar works\cite{Calandra2018,Yayev2018} on monolayer 1T-NbSe$_2$ were reported during the final stage of the preparation of this manuscript.

\appendix

\section{Density functional theory}
\label{sec:app_DFT}
We use density-functional theory  (DFT)\cite{hohenberg64_pr136_B864,kohn65_pr140_1133} to study the electronic structure of monolayer 1T-NbSe$_2$. The DFT calculations presented here are performed using the Vienna \textit{Ab-initio} Simulation Package \textsc{VASP} \cite{Kre94a, Kre99a} with the generalized gradient approximation of Perdew, Burke and Ernzerhof (PBE) \cite{perdew96_prl77_3865,perdew97_prl78_1396} for the exchange-correlation functional. 
For the structure relaxation and self-consistent electronic density and energy computation, we use an energy cutoff of $500 \eV$, and, a $\Gamma$-centered $\vec k$-mesh of size $18\times 18\times1$. 
The density of states (DOS) on the other hand is evaluated using a denser $\vec k$-mesh of size $24\times 24\times1$. 
The vacuum height ($c$) is chosen to be $20\Ang$ which is large enough to prevent interactions between periodic images. 

For the monolayer, we also study the distorted phase with $\sqrt{13}\times\sqrt{13}~$ periodic density modulation (\SoD* cluster) to account for the commensurate CDW phase observed in the STM images of 1T-NbSe$_2$ monolayers.\cite{Nakata2016}
For DFT calculations pertaining to supercells that reproduce \SoD* clusters, we choose a $\vec k$-mesh of size $6\times6\times1$. The supercell is shown in Fig.~\ref{fig:cdw_vs_undistorted}

The positions of the atoms within the unit cell for each structure are relaxed until the residual forces on them are reduced to the order of $0.01 \eV \Ang^{-1}$.
The optimized structure is further used for self-consistent electronic density and energy calculations.

\paragraph*{\textbf{Effects of non-local exchange}}

To study the effects of non-local exchange 
we use the $GW$ formalism where the quasi-particle energies -- the Kohn-Sham (KS) energies -- are renormalized by the self-energy which is evaluated as a product of the single-particle Green's function and the screened Coulomb interaction.\cite{hedin_gw65} 
In particular, we use the $G^{(0)}W^{(0)}$ approximation where the single-particle Green's function $G^{(0)}$ is the non-interacting Green's function of the KS system. 
The screened interaction $W$ is evaluated within the random-phase approximation (RPA) as
\begin{align}
\label{eq:gw_equation}
 W(\vec q,\omega) &= v_{\vec q} \left[1-\Pi(\vec q,\omega)v_{{\vec q}}\right]^{-1}, \nonumber \\
 \Pi(\vec q, \omega) &= \sum_{m, n, \vec k} |\langle \vec k-\vec q n|e^{-\I \vec q\cdot\vec r}|\vec km \rangle|^2 \nonumber \\[-3mm]
&\quad \times \frac{f_n(\vec k-\vec q)-f_m(\vec k)}{\omega+i0^++\epsilon_n(\vec k-\vec q)-\epsilon_m(\vec k)},
\end{align}
where $v_{\vec q}$ is the bare Coulomb interaction and $\Pi(q,\omega)$ is the polarizability within the RPA. 
$|\vec km \rangle$ and $\epsilon_m(\vec k)$ are the eigenvectors and eigenvalues of the KS Hamiltonian, respectively. 

Fully-converged $GW$ calculations for large supercells that account for the \SoD* clusters have enormous computational requirements of both CPU time and memory. For the supercell, we therefore use hybrid functionals that admix a certain amount of Fock exchange $ E_{{\textrm{x}}}^{\textrm{HF}}$ to the semi-local (GGA) density functional. In particular, we use the range-separated hybrid functional HSE06.\cite{hse06} 
The amount of admixing is controlled by a free parameter $\alpha$ that needs to be chosen \textit{a priori}.  

The calculations for GW and HSE06 were also performed using the Vienna \textit{Ab-initio} Simulation Package (\textsc{VASP}).\cite{Shishkin2006_vaspgw,hse06_vasp}

\section{Density functional perturbation theory}
\label{sec:app_DFPT}
The vibrational properties and thus any structural instability of a material can be inferred from its response to all possible displacements of the atoms which we calculate within the adiabatic,\cite{BorOpp27} harmonic\cite{BarKar12} and generalized gradient approximation\cite{perdew96_prl77_3865} using \textit{density-functional perturbation theory}\cite{Bar01} (DFPT). Here, in terms of the electron density $n(\vec r)$, the \textit{interatomic force constants} are given by
\begin{multline}
  C_{\vec R - \vec R'}^{\alpha \beta i j} = \frac
    {\partial^2 E}
    {\partial u_{\vec R}^{\alpha i} \, \partial u_{\vec R'}^{\beta j}} \\
  + \int \D^3 r \bigg[
    \frac{\partial V(\vec r)}{\partial u_{\vec R}^{\alpha i}}
    \frac{\partial n(\vec r)}{\partial u_{\vec R'}^{\beta j}}
    + \frac
      {\partial^2 V(\vec r)}
      {\partial u_{\vec R}^{\alpha i} \, \partial u_{\vec R'}^{\beta j}}
    n(\vec r)
  \bigg],
\end{multline}
where $E$ is the electrostatic energy of the configuration of nuclei, which imposes the external potential $V(\vec r)$ on the individual electrons, and $u_{\vec R}^{\alpha i}$ is the displacement in the $i$-th direction of the $\alpha$-th atom in the unit cell at $\vec R$. The terms in brackets account for linear and quadratic electron-phonon coupling.\cite{Nom15}

The dispersion $\omega_{\vec q}^\nu$ of the $\nu$-th phonon mode now follows from the eigenvalue equation for the \textit{dynamical matrix},
\begin{equation}
  \vec D_{\vec q} \vec e_{\vec q}^\nu = \omega_{\vec q}^\nu \vec e_{\vec q}^\nu,
  \quad
  D_{\vec q}^{\alpha \beta i j} = \sum_{\vec R}
  \frac{C_{\vec R}^{\alpha \beta i j}}{\sqrt{M_\alpha M_\beta}}
  \E^{\I \vec q \cdot \vec R},
\end{equation}
where $M_\alpha$ is a nuclear mass. Upon softening of the interatomic ``springs'' the phonon frequencies are reduced. A prevalence of negative force constants may even lead to freezing phonon modes, i.e., permanent lattice distortions, and imaginary frequencies. Such unphysical results are usually found near wave vectors $\vec q$ belonging to charge-density waves neglected in the calculation. \JB{Reference needed.}

We use \textsc{Quantum ESPRESSO}'s\cite{QE09, QE17} DFPT implementation together with ultrasoft PBE pseudopotentials from the PSLibrary 1.0.0\cite{PSLibrary} at energy cutoffs of 70\,Ry and 490\,Ry for wave functions and electron density. Monkhorst-Pack meshes\cite{MP76} of $36 \times 36 \times 1$ and $12 \times 12\times 1$ $\vec k$ and $\vec q$ points are combined with a Methfessel-Paxton smearing\cite{MP89} of 10\,mRy. Assuming a fixed unit cell height of 20\,\AA, minimizing forces and in-plane pressure to below $1 \upmu \mathrm{Ry}/\mathrm{Bohr}$ and 1\,mbar yields a lattice constant of 3.5\,\AA, in agreement with \textsc{VASP}.

\section{Estimation of screened Coulomb interaction}
\label{sec:coloumbestimation}
The relevant low energy subspace $\mathcal{C}$ consists of three bands around the Fermi level (see Fig.\ref{fig:undis_band_nbse2}) with predominantly Nb-$\lbrace d_{z^2},d_{x^2-y^2},d_{xy}\rbrace$ character, and is constructed in the Wannier basis using the \textsc{Wannier90}\cite{Mostofi2008685} package. The screened Coulomb interaction $W(\vec q,\omega)$ in Eq.~\eqref{eq:gw_equation} is evaluated in the Wannier basis, using the RPA implementation of \textsc{VASP}.\cite{Merzuk_PhD}


From the \textit{ab-initio} calculation of RPA-screened Coulomb interaction for the undistorted phase, the local Hubbard-type interaction for the flat band in the \SoD* phase is estimated according to the following scheme:

\begin{itemize}
\item We start from the static RPA-screened Coulomb matrix elements $W_{\alpha \beta \gamma \delta}(\vec q,\omega\rightarrow 0)$ for undistorted 1T-NbSe$_2$, where $\vec q$ is a wave number on a $18 \times
  18$ Monkhorst-Pack mesh and $\alpha,\beta,\gamma,\delta\in \mathcal{C}$.
\item The eigenbasis $\lbrace|e_i(\vec q)\rangle\rbrace$ of the bare Coulomb interaction  matrix $v_{\vec q}$ containing only dominating density-density terms $v_{\alpha\beta}^{\vec q}=v_{\alpha\alpha,\beta\beta}$ is used to diagonalize the corresponding screened Coulomb interaction matrix as
 \begin{equation}
 \lambda^{(i)}_{\vec q}=\langle e_i(\vec q)|W(\vec q)|e_i(\vec q)\rangle.
\end{equation}
\item We consider only the leading eigenvalue,
 \begin{equation}
    U_{\vec q} = \max_i \lambda_{\vec q}^{(i)},
  \end{equation}
  since it describes the dominating contribution to the effective interaction whenever several unit cells (as in the case of the CDW here) are involved. \cite{Malte_wfce}
\item Next, we perform a discrete Fourier transform,
  \begin{equation}
    U_{\vec R} = \frac 1 N \sum_{\vec q} \E^{\I \vec q \vec R} U_{\vec q},
  \end{equation}
  where $\vec R$ points to the unit cells in a $N = 18 \times 18$ supercell.
\item  Our estimate for the local Hubbard $U$ in the \SoD, which is formed on a $\sqrt{13} \times \sqrt{13}$ supercell, is the average of all (local and non-local) interactions within the atoms forming the \SoD
  \begin{equation}
    U = \frac 1 {N^2}\sum_{\vec R, \vec R' \in \davidsstar}U_{\vec R - \vec R'},
  \end{equation}
  where $N$ can be 7 or 13 depending on whether the outer ring of the star is excluded or not ($\vec R $ and $\vec R'$ are, however, lattice vectors of the undistorted lattice). 
\end{itemize}

\section{Dynamical Mean-Field Theory}
\label{sec:DMFT}
In order to treat correlation effects stemming from local Coulomb interactions adequately, we use the LDA++ approach,\cite{anisimov_first-principles_1997,lichtenstein__1998} i.e., we augment a DFT-based tight-binding model with local Coulomb interactions and solve the resulting generalized Hubbard model in dynamical mean-field theory\cite{georges_dynamical_1996} (DMFT). We solve the impurity problem using the continuous-time quantum Monte-Carlo algorithm (CTQMC) in the hybridization expansion\cite{EGull_ctqmc_rmp,Sagiovani_prb2012_conservedquant}. The worm algorithm\cite{worm1998,sangiovani_prb2015_worm,sangiovani_prb2016_worm} is used for the global updates in the CTQMC and five millions measurement sweeps are performed. We estimate the local Coulomb interaction $U$ as outlined in the prior Sec.~\ref{sec:coloumbestimation}. Starting from the \textit{ab-initio} solution, we perform a unitary transformation from the delocalized Kohn-Sham states to localized states, which we obtain by orthonormalizing optimized projector states available in the PAW framework implemented in \textsc{VASP}.\cite{Schulerprojector2018} Note however, that by choice of the three-band subspace here, the ``localized'' effective orbitals resulting from this projection are indeed spread over several atoms in each \SoD in the CDW supercell.


\bibliography{text}

\begin{thebibliography}{55}%
\makeatletter
\providecommand \@ifxundefined [1]{%
 \@ifx{#1\undefined}
}%
\providecommand \@ifnum [1]{%
 \ifnum #1\expandafter \@firstoftwo
 \else \expandafter \@secondoftwo
 \fi
}%
\providecommand \@ifx [1]{%
 \ifx #1\expandafter \@firstoftwo
 \else \expandafter \@secondoftwo
 \fi
}%
\providecommand \natexlab [1]{#1}%
\providecommand \enquote  [1]{``#1''}%
\providecommand \bibnamefont  [1]{#1}%
\providecommand \bibfnamefont [1]{#1}%
\providecommand \citenamefont [1]{#1}%
\providecommand \href@noop [0]{\@secondoftwo}%
\providecommand \href [0]{\begingroup \@sanitize@url \@href}%
\providecommand \@href[1]{\@@startlink{#1}\@@href}%
\providecommand \@@href[1]{\endgroup#1\@@endlink}%
\providecommand \@sanitize@url [0]{\catcode `\\12\catcode `\$12\catcode
  `\&12\catcode `\#12\catcode `\^12\catcode `\_12\catcode `\%12\relax}%
\providecommand \@@startlink[1]{}%
\providecommand \@@endlink[0]{}%
\providecommand \url  [0]{\begingroup\@sanitize@url \@url }%
\providecommand \@url [1]{\endgroup\@href {#1}{\urlprefix }}%
\providecommand \urlprefix  [0]{URL }%
\providecommand \Eprint [0]{\href }%
\providecommand \doibase [0]{http://dx.doi.org/}%
\providecommand \selectlanguage [0]{\@gobble}%
\providecommand \bibinfo  [0]{\@secondoftwo}%
\providecommand \bibfield  [0]{\@secondoftwo}%
\providecommand \translation [1]{[#1]}%
\providecommand \BibitemOpen [0]{}%
\providecommand \bibitemStop [0]{}%
\providecommand \bibitemNoStop [0]{.\EOS\space}%
\providecommand \EOS [0]{\spacefactor3000\relax}%
\providecommand \BibitemShut  [1]{\csname bibitem#1\endcsname}%
\let\auto@bib@innerbib\@empty
\bibitem [{\citenamefont {Novoselov}\ \emph {et~al.}(2005)\citenamefont
  {Novoselov}, \citenamefont {Geim}, \citenamefont {Morozov}, \citenamefont
  {Jiang}, \citenamefont {Katsnelson}, \citenamefont {Grigorieva},
  \citenamefont {Dubonos},\ and\ \citenamefont
  {Firsov}}]{novoselov_two-dimensional_2005}%
  \BibitemOpen
  \bibfield  {author} {\bibinfo {author} {\bibfnamefont {K.~S.}\ \bibnamefont
  {Novoselov}}, \bibinfo {author} {\bibfnamefont {A.~K.}\ \bibnamefont {Geim}},
  \bibinfo {author} {\bibfnamefont {S.~V.}\ \bibnamefont {Morozov}}, \bibinfo
  {author} {\bibfnamefont {D.}~\bibnamefont {Jiang}}, \bibinfo {author}
  {\bibfnamefont {M.~I.}\ \bibnamefont {Katsnelson}}, \bibinfo {author}
  {\bibfnamefont {I.~V.}\ \bibnamefont {Grigorieva}}, \bibinfo {author}
  {\bibfnamefont {S.~V.}\ \bibnamefont {Dubonos}}, \ and\ \bibinfo {author}
  {\bibfnamefont {A.~A.}\ \bibnamefont {Firsov}},\ }\href {\doibase
  10.1038/nature04233} {\bibfield  {journal} {\bibinfo  {journal} {Nature}\
  }\textbf {\bibinfo {volume} {438}},\ \bibinfo {pages} {197} (\bibinfo {year}
  {2005})}\BibitemShut {NoStop}%
\bibitem [{\citenamefont {Mak}\ \emph {et~al.}(2010)\citenamefont {Mak},
  \citenamefont {Lee}, \citenamefont {Hone}, \citenamefont {Shan},\ and\
  \citenamefont {Heinz}}]{mak_atomically_2010}%
  \BibitemOpen
  \bibfield  {author} {\bibinfo {author} {\bibfnamefont {K.~F.}\ \bibnamefont
  {Mak}}, \bibinfo {author} {\bibfnamefont {C.}~\bibnamefont {Lee}}, \bibinfo
  {author} {\bibfnamefont {J.}~\bibnamefont {Hone}}, \bibinfo {author}
  {\bibfnamefont {J.}~\bibnamefont {Shan}}, \ and\ \bibinfo {author}
  {\bibfnamefont {T.~F.}\ \bibnamefont {Heinz}},\ }\href {\doibase
  10.1103/PhysRevLett.105.136805} {\bibfield  {journal} {\bibinfo  {journal}
  {Phys. Rev. Lett.}\ }\textbf {\bibinfo {volume} {105}},\ \bibinfo {pages}
  {136805} (\bibinfo {year} {2010})}\BibitemShut {NoStop}%
\bibitem [{\citenamefont {Ge}\ \emph {et~al.}(2015)\citenamefont {Ge},
  \citenamefont {Liu}, \citenamefont {Liu}, \citenamefont {Gao}, \citenamefont
  {Qian}, \citenamefont {Xue}, \citenamefont {Liu},\ and\ \citenamefont
  {Jia}}]{ge_superconductivity_2015}%
  \BibitemOpen
  \bibfield  {author} {\bibinfo {author} {\bibfnamefont {J.-F.}\ \bibnamefont
  {Ge}}, \bibinfo {author} {\bibfnamefont {Z.-L.}\ \bibnamefont {Liu}},
  \bibinfo {author} {\bibfnamefont {C.}~\bibnamefont {Liu}}, \bibinfo {author}
  {\bibfnamefont {C.-L.}\ \bibnamefont {Gao}}, \bibinfo {author} {\bibfnamefont
  {D.}~\bibnamefont {Qian}}, \bibinfo {author} {\bibfnamefont {Q.-K.}\
  \bibnamefont {Xue}}, \bibinfo {author} {\bibfnamefont {Y.}~\bibnamefont
  {Liu}}, \ and\ \bibinfo {author} {\bibfnamefont {J.-F.}\ \bibnamefont
  {Jia}},\ }\href {\doibase 10.1038/nmat4153} {\bibfield  {journal} {\bibinfo
  {journal} {Nat Mater}\ }\textbf {\bibinfo {volume} {14}},\ \bibinfo {pages}
  {285} (\bibinfo {year} {2015})}\BibitemShut {NoStop}%
\bibitem [{\citenamefont {Xi}\ \emph {et~al.}(2015)\citenamefont {Xi},
  \citenamefont {Zhao}, \citenamefont {Wang}, \citenamefont {Berger},
  \citenamefont {Forro}, \citenamefont {Shan},\ and\ \citenamefont
  {Mak}}]{xi_strongly_2015}%
  \BibitemOpen
  \bibfield  {author} {\bibinfo {author} {\bibfnamefont {X.}~\bibnamefont
  {Xi}}, \bibinfo {author} {\bibfnamefont {L.}~\bibnamefont {Zhao}}, \bibinfo
  {author} {\bibfnamefont {Z.}~\bibnamefont {Wang}}, \bibinfo {author}
  {\bibfnamefont {H.}~\bibnamefont {Berger}}, \bibinfo {author} {\bibfnamefont
  {L.}~\bibnamefont {Forro}}, \bibinfo {author} {\bibfnamefont
  {J.}~\bibnamefont {Shan}}, \ and\ \bibinfo {author} {\bibfnamefont {K.~F.}\
  \bibnamefont {Mak}},\ }\href {\doibase 10.1038/nnano.2015.143} {\bibfield
  {journal} {\bibinfo  {journal} {Nat Nano}\ }\textbf {\bibinfo {volume}
  {10}},\ \bibinfo {pages} {765} (\bibinfo {year} {2015})}\BibitemShut
  {NoStop}%
\bibitem [{\citenamefont {Costanzo}\ \emph {et~al.}(2016)\citenamefont
  {Costanzo}, \citenamefont {Jo}, \citenamefont {Berger},\ and\ \citenamefont
  {Morpurgo}}]{costanzo_gate-induced_2016}%
  \BibitemOpen
  \bibfield  {author} {\bibinfo {author} {\bibfnamefont {D.}~\bibnamefont
  {Costanzo}}, \bibinfo {author} {\bibfnamefont {S.}~\bibnamefont {Jo}},
  \bibinfo {author} {\bibfnamefont {H.}~\bibnamefont {Berger}}, \ and\ \bibinfo
  {author} {\bibfnamefont {A.~F.}\ \bibnamefont {Morpurgo}},\ }\href {\doibase
  10.1038/nnano.2015.314} {\bibfield  {journal} {\bibinfo  {journal} {Nat
  Nano}\ }\textbf {\bibinfo {volume} {11}},\ \bibinfo {pages} {339} (\bibinfo
  {year} {2016})}\BibitemShut {NoStop}%
\bibitem [{\citenamefont {Calandra}\ \emph {et~al.}(2009)\citenamefont
  {Calandra}, \citenamefont {Mazin},\ and\ \citenamefont
  {Mauri}}]{calandra_effect_2009}%
  \BibitemOpen
  \bibfield  {author} {\bibinfo {author} {\bibfnamefont {M.}~\bibnamefont
  {Calandra}}, \bibinfo {author} {\bibfnamefont {I.~I.}\ \bibnamefont {Mazin}},
  \ and\ \bibinfo {author} {\bibfnamefont {F.}~\bibnamefont {Mauri}},\ }\href
  {\doibase 10.1103/PhysRevB.80.241108} {\bibfield  {journal} {\bibinfo
  {journal} {Phys. Rev. B}\ }\textbf {\bibinfo {volume} {80}},\ \bibinfo
  {pages} {241108} (\bibinfo {year} {2009})}\BibitemShut {NoStop}%
\bibitem [{\citenamefont {Yu}\ \emph {et~al.}(2015)\citenamefont {Yu},
  \citenamefont {Yang}, \citenamefont {Lu}, \citenamefont {Yan}, \citenamefont
  {Cho}, \citenamefont {Ma}, \citenamefont {Niu}, \citenamefont {Kim},
  \citenamefont {Son}, \citenamefont {Feng}, \citenamefont {Li}, \citenamefont
  {Cheong}, \citenamefont {Chen},\ and\ \citenamefont
  {Zhang}}]{yu_gate-tunable_2015}%
  \BibitemOpen
  \bibfield  {author} {\bibinfo {author} {\bibfnamefont {Y.}~\bibnamefont
  {Yu}}, \bibinfo {author} {\bibfnamefont {F.}~\bibnamefont {Yang}}, \bibinfo
  {author} {\bibfnamefont {X.~F.}\ \bibnamefont {Lu}}, \bibinfo {author}
  {\bibfnamefont {Y.~J.}\ \bibnamefont {Yan}}, \bibinfo {author} {\bibfnamefont
  {Y.-H.}\ \bibnamefont {Cho}}, \bibinfo {author} {\bibfnamefont
  {L.}~\bibnamefont {Ma}}, \bibinfo {author} {\bibfnamefont {X.}~\bibnamefont
  {Niu}}, \bibinfo {author} {\bibfnamefont {S.}~\bibnamefont {Kim}}, \bibinfo
  {author} {\bibfnamefont {Y.-W.}\ \bibnamefont {Son}}, \bibinfo {author}
  {\bibfnamefont {D.}~\bibnamefont {Feng}}, \bibinfo {author} {\bibfnamefont
  {S.}~\bibnamefont {Li}}, \bibinfo {author} {\bibfnamefont {S.-W.}\
  \bibnamefont {Cheong}}, \bibinfo {author} {\bibfnamefont {X.~H.}\
  \bibnamefont {Chen}}, \ and\ \bibinfo {author} {\bibfnamefont
  {Y.}~\bibnamefont {Zhang}},\ }\href {\doibase 10.1038/nnano.2014.323}
  {\bibfield  {journal} {\bibinfo  {journal} {Nat. Nano.}\ }\textbf {\bibinfo
  {volume} {10}},\ \bibinfo {pages} {270} (\bibinfo {year} {2015})}\BibitemShut
  {NoStop}%
\bibitem [{\citenamefont {Wilson}\ \emph {et~al.}(1975)\citenamefont {Wilson},
  \citenamefont {Salvo},\ and\ \citenamefont {Mahajan}}]{Wilson_Salvo_rev1975}%
  \BibitemOpen
  \bibfield  {author} {\bibinfo {author} {\bibfnamefont {J.}~\bibnamefont
  {Wilson}}, \bibinfo {author} {\bibfnamefont {F.~D.}\ \bibnamefont {Salvo}}, \
  and\ \bibinfo {author} {\bibfnamefont {S.}~\bibnamefont {Mahajan}},\ }\href
  {\doibase 10.1080/00018737500101391} {\bibfield  {journal} {\bibinfo
  {journal} {Advances in Physics}\ }\textbf {\bibinfo {volume} {24}},\ \bibinfo
  {pages} {117} (\bibinfo {year} {1975})}\BibitemShut {NoStop}%
\bibitem [{\citenamefont {Fazekas}\ and\ \citenamefont
  {Tosatti}(1979)}]{Fazekas_TaS2_79}%
  \BibitemOpen
  \bibfield  {author} {\bibinfo {author} {\bibfnamefont {P.}~\bibnamefont
  {Fazekas}}\ and\ \bibinfo {author} {\bibfnamefont {E.}~\bibnamefont
  {Tosatti}},\ }\href {\doibase 10.1080/13642817908245359} {\bibfield
  {journal} {\bibinfo  {journal} {Philosophical Magazine B}\ }\textbf {\bibinfo
  {volume} {39}},\ \bibinfo {pages} {229} (\bibinfo {year} {1979})}\BibitemShut
  {NoStop}%
\bibitem [{\citenamefont {Fazekas}\ and\ \citenamefont
  {Tosatti}(1980)}]{Fazekas_TaS2_80}%
  \BibitemOpen
  \bibfield  {author} {\bibinfo {author} {\bibfnamefont {P.}~\bibnamefont
  {Fazekas}}\ and\ \bibinfo {author} {\bibfnamefont {E.}~\bibnamefont
  {Tosatti}},\ }\href {\doibase 10.1016/0378-4363(80)90229-6} {\bibfield
  {journal} {\bibinfo  {journal} {Physica B: Physics of Condensed Matter \& C:
  Atomic, Molecular and Plasma Physics, Optics}\ }\textbf {\bibinfo {volume}
  {99}},\ \bibinfo {pages} {183} (\bibinfo {year} {1980})}\BibitemShut
  {NoStop}%
\bibitem [{\citenamefont {Albertini}\ \emph {et~al.}(2016)\citenamefont
  {Albertini}, \citenamefont {Zhao}, \citenamefont {McCann}, \citenamefont
  {Feng}, \citenamefont {Terrones}, \citenamefont {Freericks}, \citenamefont
  {Robinson},\ and\ \citenamefont {Liu}}]{albertini_zone-center_2016}%
  \BibitemOpen
  \bibfield  {author} {\bibinfo {author} {\bibfnamefont {O.~R.}\ \bibnamefont
  {Albertini}}, \bibinfo {author} {\bibfnamefont {R.}~\bibnamefont {Zhao}},
  \bibinfo {author} {\bibfnamefont {R.~L.}\ \bibnamefont {McCann}}, \bibinfo
  {author} {\bibfnamefont {S.}~\bibnamefont {Feng}}, \bibinfo {author}
  {\bibfnamefont {M.}~\bibnamefont {Terrones}}, \bibinfo {author}
  {\bibfnamefont {J.~K.}\ \bibnamefont {Freericks}}, \bibinfo {author}
  {\bibfnamefont {J.~A.}\ \bibnamefont {Robinson}}, \ and\ \bibinfo {author}
  {\bibfnamefont {A.~Y.}\ \bibnamefont {Liu}},\ }\href {\doibase
  10.1103/PhysRevB.93.214109} {\bibfield  {journal} {\bibinfo  {journal} {Phys.
  Rev. B}\ }\textbf {\bibinfo {volume} {93}},\ \bibinfo {pages} {214109}
  (\bibinfo {year} {2016})}\BibitemShut {NoStop}%
\bibitem [{\citenamefont {Darancet}\ \emph {et~al.}(2014)\citenamefont
  {Darancet}, \citenamefont {Millis},\ and\ \citenamefont
  {Marianetti}}]{Pierre2014}%
  \BibitemOpen
  \bibfield  {author} {\bibinfo {author} {\bibfnamefont {P.}~\bibnamefont
  {Darancet}}, \bibinfo {author} {\bibfnamefont {A.~J.}\ \bibnamefont
  {Millis}}, \ and\ \bibinfo {author} {\bibfnamefont {C.~A.}\ \bibnamefont
  {Marianetti}},\ }\href {\doibase 10.1103/PhysRevB.90.045134} {\bibfield
  {journal} {\bibinfo  {journal} {Phys. Rev. B}\ }\textbf {\bibinfo {volume}
  {90}},\ \bibinfo {pages} {045134} (\bibinfo {year} {2014})}\BibitemShut
  {NoStop}%
\bibitem [{\citenamefont {Nakata}\ \emph {et~al.}(2016)\citenamefont {Nakata},
  \citenamefont {Sugawara}, \citenamefont {Shimizu}, \citenamefont {Okada},
  \citenamefont {Han}, \citenamefont {Hitosugi}, \citenamefont {Ueno},
  \citenamefont {Sato},\ and\ \citenamefont {Takahashi}}]{Nakata2016}%
  \BibitemOpen
  \bibfield  {author} {\bibinfo {author} {\bibfnamefont {Y.}~\bibnamefont
  {Nakata}}, \bibinfo {author} {\bibfnamefont {K.}~\bibnamefont {Sugawara}},
  \bibinfo {author} {\bibfnamefont {R.}~\bibnamefont {Shimizu}}, \bibinfo
  {author} {\bibfnamefont {Y.}~\bibnamefont {Okada}}, \bibinfo {author}
  {\bibfnamefont {P.}~\bibnamefont {Han}}, \bibinfo {author} {\bibfnamefont
  {T.}~\bibnamefont {Hitosugi}}, \bibinfo {author} {\bibfnamefont
  {K.}~\bibnamefont {Ueno}}, \bibinfo {author} {\bibfnamefont {T.}~\bibnamefont
  {Sato}}, \ and\ \bibinfo {author} {\bibfnamefont {T.}~\bibnamefont
  {Takahashi}},\ }\href {http://dx.doi.org/10.1038/am.2016.157} {\bibfield
  {journal} {\bibinfo  {journal} {NPG Asia Mater}\ }\textbf {\bibinfo {volume}
  {8}},\ \bibinfo {pages} {e321} (\bibinfo {year} {2016})}\BibitemShut
  {NoStop}%
\bibitem [{\citenamefont {Ugeda}\ \emph {et~al.}(2016)\citenamefont {Ugeda},
  \citenamefont {Bradley}, \citenamefont {Zhang}, \citenamefont {Onishi},
  \citenamefont {Chen}, \citenamefont {Ruan}, \citenamefont
  {Ojeda-Aristizabal}, \citenamefont {Ryu}, \citenamefont {Edmonds},
  \citenamefont {Tsai}, \citenamefont {Riss}, \citenamefont {Mo}, \citenamefont
  {Lee}, \citenamefont {Zettl}, \citenamefont {Hussain}, \citenamefont {Shen},\
  and\ \citenamefont {Crommie}}]{ugeda_characterization_2016}%
  \BibitemOpen
  \bibfield  {author} {\bibinfo {author} {\bibfnamefont {M.~M.}\ \bibnamefont
  {Ugeda}}, \bibinfo {author} {\bibfnamefont {A.~J.}\ \bibnamefont {Bradley}},
  \bibinfo {author} {\bibfnamefont {Y.}~\bibnamefont {Zhang}}, \bibinfo
  {author} {\bibfnamefont {S.}~\bibnamefont {Onishi}}, \bibinfo {author}
  {\bibfnamefont {Y.}~\bibnamefont {Chen}}, \bibinfo {author} {\bibfnamefont
  {W.}~\bibnamefont {Ruan}}, \bibinfo {author} {\bibfnamefont {C.}~\bibnamefont
  {Ojeda-Aristizabal}}, \bibinfo {author} {\bibfnamefont {H.}~\bibnamefont
  {Ryu}}, \bibinfo {author} {\bibfnamefont {M.~T.}\ \bibnamefont {Edmonds}},
  \bibinfo {author} {\bibfnamefont {H.-Z.}\ \bibnamefont {Tsai}}, \bibinfo
  {author} {\bibfnamefont {A.}~\bibnamefont {Riss}}, \bibinfo {author}
  {\bibfnamefont {S.-K.}\ \bibnamefont {Mo}}, \bibinfo {author} {\bibfnamefont
  {D.}~\bibnamefont {Lee}}, \bibinfo {author} {\bibfnamefont {A.}~\bibnamefont
  {Zettl}}, \bibinfo {author} {\bibfnamefont {Z.}~\bibnamefont {Hussain}},
  \bibinfo {author} {\bibfnamefont {Z.-X.}\ \bibnamefont {Shen}}, \ and\
  \bibinfo {author} {\bibfnamefont {M.~F.}\ \bibnamefont {Crommie}},\ }\href
  {\doibase 10.1038/nphys3527} {\bibfield  {journal} {\bibinfo  {journal} {Nat
  Phys}\ }\textbf {\bibinfo {volume} {12}},\ \bibinfo {pages} {92} (\bibinfo
  {year} {2016})}\BibitemShut {NoStop}%
\bibitem [{\citenamefont {Ugeda}\ \emph {et~al.}(2015)\citenamefont {Ugeda},
  \citenamefont {Bradley}, \citenamefont {Zhang}, \citenamefont {Onishi},
  \citenamefont {Chen}, \citenamefont {Ruan}, \citenamefont
  {Ojeda-Aristizabal}, \citenamefont {Ryu}, \citenamefont {Edmonds},
  \citenamefont {Tsai}, \citenamefont {Riss}, \citenamefont {Mo}, \citenamefont
  {Lee}, \citenamefont {Zettl}, \citenamefont {Hussain}, \citenamefont {Shen},\
  and\ \citenamefont {Crommie}}]{Uge15}%
  \BibitemOpen
  \bibfield  {author} {\bibinfo {author} {\bibfnamefont {M.~M.}\ \bibnamefont
  {Ugeda}}, \bibinfo {author} {\bibfnamefont {A.~J.}\ \bibnamefont {Bradley}},
  \bibinfo {author} {\bibfnamefont {Y.}~\bibnamefont {Zhang}}, \bibinfo
  {author} {\bibfnamefont {S.}~\bibnamefont {Onishi}}, \bibinfo {author}
  {\bibfnamefont {Y.}~\bibnamefont {Chen}}, \bibinfo {author} {\bibfnamefont
  {W.}~\bibnamefont {Ruan}}, \bibinfo {author} {\bibfnamefont {C.}~\bibnamefont
  {Ojeda-Aristizabal}}, \bibinfo {author} {\bibfnamefont {H.}~\bibnamefont
  {Ryu}}, \bibinfo {author} {\bibfnamefont {M.~T.}\ \bibnamefont {Edmonds}},
  \bibinfo {author} {\bibfnamefont {H.-Z.}\ \bibnamefont {Tsai}}, \bibinfo
  {author} {\bibfnamefont {A.}~\bibnamefont {Riss}}, \bibinfo {author}
  {\bibfnamefont {S.-K.}\ \bibnamefont {Mo}}, \bibinfo {author} {\bibfnamefont
  {D.}~\bibnamefont {Lee}}, \bibinfo {author} {\bibfnamefont {A.}~\bibnamefont
  {Zettl}}, \bibinfo {author} {\bibfnamefont {Z.}~\bibnamefont {Hussain}},
  \bibinfo {author} {\bibfnamefont {Z.-X.}\ \bibnamefont {Shen}}, \ and\
  \bibinfo {author} {\bibfnamefont {M.~F.}\ \bibnamefont {Crommie}},\ }\href
  {\doibase 10.1038/nphys3527} {\bibfield  {journal} {\bibinfo  {journal}
  {Nature Physics}\ }\textbf {\bibinfo {volume} {12}},\ \bibinfo {pages} {92}
  (\bibinfo {year} {2015})}\BibitemShut {NoStop}%
\bibitem [{\citenamefont {Medeiros}\ \emph {et~al.}(2014)\citenamefont
  {Medeiros}, \citenamefont {Stafstr\"om},\ and\ \citenamefont
  {Bj\"ork}}]{bandup1}%
  \BibitemOpen
  \bibfield  {author} {\bibinfo {author} {\bibfnamefont {P.~V.~C.}\
  \bibnamefont {Medeiros}}, \bibinfo {author} {\bibfnamefont {S.}~\bibnamefont
  {Stafstr\"om}}, \ and\ \bibinfo {author} {\bibfnamefont {J.}~\bibnamefont
  {Bj\"ork}},\ }\href {\doibase 10.1103/PhysRevB.89.041407} {\bibfield
  {journal} {\bibinfo  {journal} {Phys. Rev. B}\ }\textbf {\bibinfo {volume}
  {89}},\ \bibinfo {pages} {041407} (\bibinfo {year} {2014})}\BibitemShut
  {NoStop}%
\bibitem [{\citenamefont {Medeiros}\ \emph {et~al.}(2015)\citenamefont
  {Medeiros}, \citenamefont {Tsirkin}, \citenamefont {Stafstr\"om},\ and\
  \citenamefont {Bj\"ork}}]{bandup2}%
  \BibitemOpen
  \bibfield  {author} {\bibinfo {author} {\bibfnamefont {P.~V.~C.}\
  \bibnamefont {Medeiros}}, \bibinfo {author} {\bibfnamefont {S.~S.}\
  \bibnamefont {Tsirkin}}, \bibinfo {author} {\bibfnamefont {S.}~\bibnamefont
  {Stafstr\"om}}, \ and\ \bibinfo {author} {\bibfnamefont {J.}~\bibnamefont
  {Bj\"ork}},\ }\href {\doibase 10.1103/PhysRevB.91.041116} {\bibfield
  {journal} {\bibinfo  {journal} {Phys. Rev. B}\ }\textbf {\bibinfo {volume}
  {91}},\ \bibinfo {pages} {041116} (\bibinfo {year} {2015})}\BibitemShut
  {NoStop}%
\bibitem [{\citenamefont {Czyżyk}\ and\ \citenamefont
  {Sawatzky}(1994)}]{czyzyk_local-density_1994}%
  \BibitemOpen
  \bibfield  {author} {\bibinfo {author} {\bibfnamefont {M.~T.}\ \bibnamefont
  {Czyżyk}}\ and\ \bibinfo {author} {\bibfnamefont {G.~A.}\ \bibnamefont
  {Sawatzky}},\ }\href {\doibase 10.1103/PhysRevB.49.14211} {\bibfield
  {journal} {\bibinfo  {journal} {Physical Review B}\ }\textbf {\bibinfo
  {volume} {49}},\ \bibinfo {pages} {14211} (\bibinfo {year}
  {1994})}\BibitemShut {NoStop}%
\bibitem [{\citenamefont {Parragh}\ \emph
  {et~al.}(2012{\natexlab{a}})\citenamefont {Parragh}, \citenamefont {Toschi},
  \citenamefont {Held},\ and\ \citenamefont
  {Sangiovanni}}]{parragh_conserved_2012}%
  \BibitemOpen
  \bibfield  {author} {\bibinfo {author} {\bibfnamefont {N.}~\bibnamefont
  {Parragh}}, \bibinfo {author} {\bibfnamefont {A.}~\bibnamefont {Toschi}},
  \bibinfo {author} {\bibfnamefont {K.}~\bibnamefont {Held}}, \ and\ \bibinfo
  {author} {\bibfnamefont {G.}~\bibnamefont {Sangiovanni}},\ }\href {\doibase
  10.1103/PhysRevB.86.155158} {\bibfield  {journal} {\bibinfo  {journal} {Phys.
  Rev. B}\ }\textbf {\bibinfo {volume} {86}},\ \bibinfo {pages} {155158}
  (\bibinfo {year} {2012}{\natexlab{a}})}\BibitemShut {NoStop}%
\bibitem [{\citenamefont {Wallerberger}\ \emph {et~al.}(2018)\citenamefont
  {Wallerberger}, \citenamefont {Hausoel}, \citenamefont {Gunacker},
  \citenamefont {Kowalski}, \citenamefont {Parragh}, \citenamefont {Goth},
  \citenamefont {Held},\ and\ \citenamefont
  {Sangiovanni}}]{wallerberger_w2dynamics:_2018}%
  \BibitemOpen
  \bibfield  {author} {\bibinfo {author} {\bibfnamefont {M.}~\bibnamefont
  {Wallerberger}}, \bibinfo {author} {\bibfnamefont {A.}~\bibnamefont
  {Hausoel}}, \bibinfo {author} {\bibfnamefont {P.}~\bibnamefont {Gunacker}},
  \bibinfo {author} {\bibfnamefont {A.}~\bibnamefont {Kowalski}}, \bibinfo
  {author} {\bibfnamefont {N.}~\bibnamefont {Parragh}}, \bibinfo {author}
  {\bibfnamefont {F.}~\bibnamefont {Goth}}, \bibinfo {author} {\bibfnamefont
  {K.}~\bibnamefont {Held}}, \ and\ \bibinfo {author} {\bibfnamefont
  {G.}~\bibnamefont {Sangiovanni}},\ }\href {http://arxiv.org/abs/1801.10209}
  {\bibfield  {journal} {\bibinfo  {journal} {arXiv:1801.10209 [cond-mat,
  physics:physics]}\ } (\bibinfo {year} {2018})}\BibitemShut {NoStop}%
\bibitem [{\citenamefont {Perfetti}\ \emph {et~al.}(2006)\citenamefont
  {Perfetti}, \citenamefont {Loukakos}, \citenamefont {Lisowski}, \citenamefont
  {Bovensiepen}, \citenamefont {Berger}, \citenamefont {Biermann},
  \citenamefont {Cornaglia}, \citenamefont {Georges},\ and\ \citenamefont
  {Wolf}}]{Perfetti_TaS2_2006}%
  \BibitemOpen
  \bibfield  {author} {\bibinfo {author} {\bibfnamefont {L.}~\bibnamefont
  {Perfetti}}, \bibinfo {author} {\bibfnamefont {P.~A.}\ \bibnamefont
  {Loukakos}}, \bibinfo {author} {\bibfnamefont {M.}~\bibnamefont {Lisowski}},
  \bibinfo {author} {\bibfnamefont {U.}~\bibnamefont {Bovensiepen}}, \bibinfo
  {author} {\bibfnamefont {H.}~\bibnamefont {Berger}}, \bibinfo {author}
  {\bibfnamefont {S.}~\bibnamefont {Biermann}}, \bibinfo {author}
  {\bibfnamefont {P.~S.}\ \bibnamefont {Cornaglia}}, \bibinfo {author}
  {\bibfnamefont {A.}~\bibnamefont {Georges}}, \ and\ \bibinfo {author}
  {\bibfnamefont {M.}~\bibnamefont {Wolf}},\ }\href {\doibase
  10.1103/PhysRevLett.97.067402} {\bibfield  {journal} {\bibinfo  {journal}
  {Phys. Rev. Lett.}\ }\textbf {\bibinfo {volume} {97}},\ \bibinfo {pages}
  {067402} (\bibinfo {year} {2006})}\BibitemShut {NoStop}%
\bibitem [{\citenamefont {Freericks}\ \emph {et~al.}()\citenamefont
  {Freericks}, \citenamefont {Krishnamurthy}, \citenamefont {Ge}, \citenamefont
  {Liu},\ and\ \citenamefont {Pruschke}}]{Freericks_pruschke_tas_2009}%
  \BibitemOpen
  \bibfield  {author} {\bibinfo {author} {\bibfnamefont {J.~K.}\ \bibnamefont
  {Freericks}}, \bibinfo {author} {\bibfnamefont {H.~R.}\ \bibnamefont
  {Krishnamurthy}}, \bibinfo {author} {\bibfnamefont {Y.}~\bibnamefont {Ge}},
  \bibinfo {author} {\bibfnamefont {A.~Y.}\ \bibnamefont {Liu}}, \ and\
  \bibinfo {author} {\bibfnamefont {T.}~\bibnamefont {Pruschke}},\ }\href
  {\doibase 10.1002/pssb.200881555} {\bibfield  {journal} {\bibinfo  {journal}
  {physica status solidi (b)}\ }\textbf {\bibinfo {volume} {246}},\ \bibinfo
  {pages} {948}}\BibitemShut {NoStop}%
\bibitem [{\citenamefont {{Calandra}}(2018)}]{Calandra2018}%
  \BibitemOpen
  \bibfield  {author} {\bibinfo {author} {\bibfnamefont {M.}~\bibnamefont
  {{Calandra}}},\ }\href@noop {} {\bibfield  {journal} {\bibinfo  {journal}
  {ArXiv e-prints}\ } (\bibinfo {year} {2018})},\ \Eprint
  {http://arxiv.org/abs/1803.08361} {arXiv:1803.08361 [cond-mat.mtrl-sci]}
  \BibitemShut {NoStop}%
\bibitem [{\citenamefont {{Pasquier}}\ and\ \citenamefont
  {{Yayev}}(2018)}]{Yayev2018}%
  \BibitemOpen
  \bibfield  {author} {\bibinfo {author} {\bibfnamefont {D.}~\bibnamefont
  {{Pasquier}}}\ and\ \bibinfo {author} {\bibfnamefont {O.~V.}\ \bibnamefont
  {{Yayev}}},\ }\href@noop {} {\bibfield  {journal} {\bibinfo  {journal} {ArXiv
  e-prints}\ } (\bibinfo {year} {2018})},\ \Eprint
  {http://arxiv.org/abs/1803.10727} {arXiv:1803.10727 [cond-mat.str-el]}
  \BibitemShut {NoStop}%
\bibitem [{\citenamefont {Hohenberg}\ and\ \citenamefont
  {Kohn}(1964)}]{hohenberg64_pr136_B864}%
  \BibitemOpen
  \bibfield  {author} {\bibinfo {author} {\bibfnamefont {P.}~\bibnamefont
  {Hohenberg}}\ and\ \bibinfo {author} {\bibfnamefont {W.}~\bibnamefont
  {Kohn}},\ }\href {\doibase 10.1103/PhysRev.136.B864} {\bibfield  {journal}
  {\bibinfo  {journal} {Phys. Rev.}\ }\textbf {\bibinfo {volume} {136}},\
  \bibinfo {pages} {B864} (\bibinfo {year} {1964})}\BibitemShut {NoStop}%
\bibitem [{\citenamefont {Kohn}\ and\ \citenamefont
  {Sham}(1965)}]{kohn65_pr140_1133}%
  \BibitemOpen
  \bibfield  {author} {\bibinfo {author} {\bibfnamefont {W.}~\bibnamefont
  {Kohn}}\ and\ \bibinfo {author} {\bibfnamefont {L.~J.}\ \bibnamefont
  {Sham}},\ }\href {\doibase 10.1103/PhysRev.140.A1133} {\bibfield  {journal}
  {\bibinfo  {journal} {Phys. Rev.}\ }\textbf {\bibinfo {volume} {140}},\
  \bibinfo {pages} {A1133} (\bibinfo {year} {1965})}\BibitemShut {NoStop}%
\bibitem [{\citenamefont {Kresse}\ and\ \citenamefont {Hafner}(1994)}]{Kre94a}%
  \BibitemOpen
  \bibfield  {author} {\bibinfo {author} {\bibfnamefont {G.}~\bibnamefont
  {Kresse}}\ and\ \bibinfo {author} {\bibfnamefont {J.}~\bibnamefont
  {Hafner}},\ }\href {http://stacks.iop.org/0953-8984/6/i=40/a=015} {\bibfield
  {journal} {\bibinfo  {journal} {Journal of Physics: Condensed Matter}\
  }\textbf {\bibinfo {volume} {6}},\ \bibinfo {pages} {8245} (\bibinfo {year}
  {1994})}\BibitemShut {NoStop}%
\bibitem [{\citenamefont {Kresse}\ and\ \citenamefont
  {Joubert}(1999)}]{Kre99a}%
  \BibitemOpen
  \bibfield  {author} {\bibinfo {author} {\bibfnamefont {G.}~\bibnamefont
  {Kresse}}\ and\ \bibinfo {author} {\bibfnamefont {D.}~\bibnamefont
  {Joubert}},\ }\href {\doibase 10.1103/PhysRevB.59.1758} {\bibfield  {journal}
  {\bibinfo  {journal} {Phys. Rev. B}\ }\textbf {\bibinfo {volume} {59}},\
  \bibinfo {pages} {1758} (\bibinfo {year} {1999})}\BibitemShut {NoStop}%
\bibitem [{\citenamefont {Perdew}\ \emph {et~al.}(1996)\citenamefont {Perdew},
  \citenamefont {Burke},\ and\ \citenamefont
  {Ernzerhof}}]{perdew96_prl77_3865}%
  \BibitemOpen
  \bibfield  {author} {\bibinfo {author} {\bibfnamefont {J.~P.}\ \bibnamefont
  {Perdew}}, \bibinfo {author} {\bibfnamefont {K.}~\bibnamefont {Burke}}, \
  and\ \bibinfo {author} {\bibfnamefont {M.}~\bibnamefont {Ernzerhof}},\
  }\href@noop {} {\bibfield  {journal} {\bibinfo  {journal} {Phys. Rev. Lett}\
  }\textbf {\bibinfo {volume} {77}},\ \bibinfo {pages} {3865} (\bibinfo {year}
  {1996})}\BibitemShut {NoStop}%
\bibitem [{\citenamefont {Perdew}\ \emph {et~al.}(1997)\citenamefont {Perdew},
  \citenamefont {Burke},\ and\ \citenamefont
  {Ernzerhof}}]{perdew97_prl78_1396}%
  \BibitemOpen
  \bibfield  {author} {\bibinfo {author} {\bibfnamefont {J.}~\bibnamefont
  {Perdew}}, \bibinfo {author} {\bibfnamefont {K.}~\bibnamefont {Burke}}, \
  and\ \bibinfo {author} {\bibfnamefont {M.}~\bibnamefont {Ernzerhof}},\
  }\href@noop {} {\bibfield  {journal} {\bibinfo  {journal} {Phys. Rev. Lett}\
  }\textbf {\bibinfo {volume} {78}},\ \bibinfo {pages} {1396} (\bibinfo {year}
  {1997})}\BibitemShut {NoStop}%
\bibitem [{\citenamefont {Hedin}(1965)}]{hedin_gw65}%
  \BibitemOpen
  \bibfield  {author} {\bibinfo {author} {\bibfnamefont {L.}~\bibnamefont
  {Hedin}},\ }\href {\doibase 10.1103/PhysRev.139.A796} {\bibfield  {journal}
  {\bibinfo  {journal} {Phys. Rev.}\ }\textbf {\bibinfo {volume} {139}},\
  \bibinfo {pages} {A796} (\bibinfo {year} {1965})}\BibitemShut {NoStop}%
\bibitem [{\citenamefont {Krukau}\ \emph {et~al.}(2006)\citenamefont {Krukau},
  \citenamefont {Vydrov}, \citenamefont {Izmaylov},\ and\ \citenamefont
  {Scuseria}}]{hse06}%
  \BibitemOpen
  \bibfield  {author} {\bibinfo {author} {\bibfnamefont {A.~V.}\ \bibnamefont
  {Krukau}}, \bibinfo {author} {\bibfnamefont {O.~A.}\ \bibnamefont {Vydrov}},
  \bibinfo {author} {\bibfnamefont {A.~F.}\ \bibnamefont {Izmaylov}}, \ and\
  \bibinfo {author} {\bibfnamefont {G.~E.}\ \bibnamefont {Scuseria}},\ }\href
  {\doibase 10.1063/1.2404663} {\bibfield  {journal} {\bibinfo  {journal} {The
  Journal of Chemical Physics}\ }\textbf {\bibinfo {volume} {125}},\ \bibinfo
  {pages} {224106} (\bibinfo {year} {2006})},\ \Eprint
  {http://arxiv.org/abs/http://dx.doi.org/10.1063/1.2404663}
  {http://dx.doi.org/10.1063/1.2404663} \BibitemShut {NoStop}%
\bibitem [{\citenamefont {Shishkin}\ and\ \citenamefont
  {Kresse}(2006)}]{Shishkin2006_vaspgw}%
  \BibitemOpen
  \bibfield  {author} {\bibinfo {author} {\bibfnamefont {M.}~\bibnamefont
  {Shishkin}}\ and\ \bibinfo {author} {\bibfnamefont {G.}~\bibnamefont
  {Kresse}},\ }\href {\doibase 10.1103/PhysRevB.74.035101} {\bibfield
  {journal} {\bibinfo  {journal} {Phys. Rev. B}\ }\textbf {\bibinfo {volume}
  {74}},\ \bibinfo {pages} {035101} (\bibinfo {year} {2006})}\BibitemShut
  {NoStop}%
\bibitem [{\citenamefont {Paier}\ \emph {et~al.}(2005)\citenamefont {Paier},
  \citenamefont {Hirschl}, \citenamefont {Marsman},\ and\ \citenamefont
  {Kresse}}]{hse06_vasp}%
  \BibitemOpen
  \bibfield  {author} {\bibinfo {author} {\bibfnamefont {J.}~\bibnamefont
  {Paier}}, \bibinfo {author} {\bibfnamefont {R.}~\bibnamefont {Hirschl}},
  \bibinfo {author} {\bibfnamefont {M.}~\bibnamefont {Marsman}}, \ and\
  \bibinfo {author} {\bibfnamefont {G.}~\bibnamefont {Kresse}},\ }\href
  {\doibase 10.1063/1.1926272} {\bibfield  {journal} {\bibinfo  {journal} {The
  Journal of Chemical Physics}\ }\textbf {\bibinfo {volume} {122}},\ \bibinfo
  {pages} {234102} (\bibinfo {year} {2005})},\ \Eprint
  {http://arxiv.org/abs/https://doi.org/10.1063/1.1926272}
  {https://doi.org/10.1063/1.1926272} \BibitemShut {NoStop}%
\bibitem [{\citenamefont {Born}\ and\ \citenamefont
  {Oppenheimer}(1927)}]{BorOpp27}%
  \BibitemOpen
  \bibfield  {author} {\bibinfo {author} {\bibfnamefont {M.}~\bibnamefont
  {Born}}\ and\ \bibinfo {author} {\bibfnamefont {J.~R.}\ \bibnamefont
  {Oppenheimer}},\ }\href {\doibase 10.1002/andp.19273892002} {\bibfield
  {journal} {\bibinfo  {journal} {Annalen der Physik}\ }\textbf {\bibinfo
  {volume} {389}} (\bibinfo {year} {1927}),\
  10.1002/andp.19273892002}\BibitemShut {NoStop}%
\bibitem [{\citenamefont {von K\'arm\'an}\ and\ \citenamefont
  {Born}(1912)}]{BarKar12}%
  \BibitemOpen
  \bibfield  {author} {\bibinfo {author} {\bibfnamefont {T.}~\bibnamefont {von
  K\'arm\'an}}\ and\ \bibinfo {author} {\bibfnamefont {M.}~\bibnamefont
  {Born}},\ }\href@noop {} {\bibfield  {journal} {\bibinfo  {journal} {Physik.
  Zeitschr.}\ }\textbf {\bibinfo {volume} {13}},\ \bibinfo {pages} {297}
  (\bibinfo {year} {1912})}\BibitemShut {NoStop}%
\bibitem [{\citenamefont {Baroni}\ \emph {et~al.}(2001)\citenamefont {Baroni},
  \citenamefont {de~Gironcoli}, \citenamefont {Dal~Corso},\ and\ \citenamefont
  {Giannozzi}}]{Bar01}%
  \BibitemOpen
  \bibfield  {author} {\bibinfo {author} {\bibfnamefont {S.}~\bibnamefont
  {Baroni}}, \bibinfo {author} {\bibfnamefont {S.}~\bibnamefont
  {de~Gironcoli}}, \bibinfo {author} {\bibfnamefont {A.}~\bibnamefont
  {Dal~Corso}}, \ and\ \bibinfo {author} {\bibfnamefont {P.}~\bibnamefont
  {Giannozzi}},\ }\href {\doibase 10.1103/RevModPhys.73.515} {\bibfield
  {journal} {\bibinfo  {journal} {Rev. Mod. Phys.}\ }\textbf {\bibinfo {volume}
  {73}},\ \bibinfo {pages} {515} (\bibinfo {year} {2001})}\BibitemShut
  {NoStop}%
\bibitem [{\citenamefont {Nomura}\ and\ \citenamefont {Arita}(2015)}]{Nom15}%
  \BibitemOpen
  \bibfield  {author} {\bibinfo {author} {\bibfnamefont {Y.}~\bibnamefont
  {Nomura}}\ and\ \bibinfo {author} {\bibfnamefont {R.}~\bibnamefont {Arita}},\
  }\href {\doibase 10.1103/PhysRevB.92.245108} {\bibfield  {journal} {\bibinfo
  {journal} {Phys. Rev. B}\ }\textbf {\bibinfo {volume} {92}},\ \bibinfo
  {pages} {245108} (\bibinfo {year} {2015})}\BibitemShut {NoStop}%
\bibitem [{\citenamefont {Giannozzi}\ \emph {et~al.}(2009)\citenamefont
  {Giannozzi}, \citenamefont {Baroni}, \citenamefont {Bonini}, \citenamefont
  {Calandra}, \citenamefont {Car}, \citenamefont {Cavazzoni}, \citenamefont
  {Ceresoli}, \citenamefont {Chiarotti}, \citenamefont {Cococcioni},
  \citenamefont {Dabo}, \citenamefont {{Dal Corso}}, \citenamefont
  {de~Gironcoli}, \citenamefont {Fabris}, \citenamefont {Fratesi},
  \citenamefont {Gebauer}, \citenamefont {Gerstmann}, \citenamefont
  {Gougoussis}, \citenamefont {Kokalj}, \citenamefont {Lazzeri}, \citenamefont
  {Martin-Samos}, \citenamefont {Marzari}, \citenamefont {Mauri}, \citenamefont
  {Mazzarello}, \citenamefont {Paolini}, \citenamefont {Pasquarello},
  \citenamefont {Paulatto}, \citenamefont {Sbraccia}, \citenamefont {Scandolo},
  \citenamefont {Sclauzero}, \citenamefont {Seitsonen}, \citenamefont
  {Smogunov}, \citenamefont {Umari},\ and\ \citenamefont
  {Wentzcovitch}}]{QE09}%
  \BibitemOpen
  \bibfield  {author} {\bibinfo {author} {\bibfnamefont {P.}~\bibnamefont
  {Giannozzi}}, \bibinfo {author} {\bibfnamefont {S.}~\bibnamefont {Baroni}},
  \bibinfo {author} {\bibfnamefont {N.}~\bibnamefont {Bonini}}, \bibinfo
  {author} {\bibfnamefont {M.}~\bibnamefont {Calandra}}, \bibinfo {author}
  {\bibfnamefont {R.}~\bibnamefont {Car}}, \bibinfo {author} {\bibfnamefont
  {C.}~\bibnamefont {Cavazzoni}}, \bibinfo {author} {\bibfnamefont
  {D.}~\bibnamefont {Ceresoli}}, \bibinfo {author} {\bibfnamefont {G.~L.}\
  \bibnamefont {Chiarotti}}, \bibinfo {author} {\bibfnamefont {M.}~\bibnamefont
  {Cococcioni}}, \bibinfo {author} {\bibfnamefont {I.}~\bibnamefont {Dabo}},
  \bibinfo {author} {\bibfnamefont {A.}~\bibnamefont {{Dal Corso}}}, \bibinfo
  {author} {\bibfnamefont {S.}~\bibnamefont {de~Gironcoli}}, \bibinfo {author}
  {\bibfnamefont {S.}~\bibnamefont {Fabris}}, \bibinfo {author} {\bibfnamefont
  {G.}~\bibnamefont {Fratesi}}, \bibinfo {author} {\bibfnamefont
  {R.}~\bibnamefont {Gebauer}}, \bibinfo {author} {\bibfnamefont
  {U.}~\bibnamefont {Gerstmann}}, \bibinfo {author} {\bibfnamefont
  {C.}~\bibnamefont {Gougoussis}}, \bibinfo {author} {\bibfnamefont
  {A.}~\bibnamefont {Kokalj}}, \bibinfo {author} {\bibfnamefont
  {M.}~\bibnamefont {Lazzeri}}, \bibinfo {author} {\bibfnamefont
  {L.}~\bibnamefont {Martin-Samos}}, \bibinfo {author} {\bibfnamefont
  {N.}~\bibnamefont {Marzari}}, \bibinfo {author} {\bibfnamefont
  {F.}~\bibnamefont {Mauri}}, \bibinfo {author} {\bibfnamefont
  {R.}~\bibnamefont {Mazzarello}}, \bibinfo {author} {\bibfnamefont
  {S.}~\bibnamefont {Paolini}}, \bibinfo {author} {\bibfnamefont
  {A.}~\bibnamefont {Pasquarello}}, \bibinfo {author} {\bibfnamefont
  {L.}~\bibnamefont {Paulatto}}, \bibinfo {author} {\bibfnamefont
  {C.}~\bibnamefont {Sbraccia}}, \bibinfo {author} {\bibfnamefont
  {S.}~\bibnamefont {Scandolo}}, \bibinfo {author} {\bibfnamefont
  {G.}~\bibnamefont {Sclauzero}}, \bibinfo {author} {\bibfnamefont {A.~P.}\
  \bibnamefont {Seitsonen}}, \bibinfo {author} {\bibfnamefont {A.}~\bibnamefont
  {Smogunov}}, \bibinfo {author} {\bibfnamefont {P.}~\bibnamefont {Umari}}, \
  and\ \bibinfo {author} {\bibfnamefont {R.~M.}\ \bibnamefont {Wentzcovitch}},\
  }\href {http://www.quantum-espresso.org} {\bibfield  {journal} {\bibinfo
  {journal} {Journal of Physics: Condensed Matter}\ }\textbf {\bibinfo {volume}
  {21}},\ \bibinfo {pages} {395502} (\bibinfo {year} {2009})}\BibitemShut
  {NoStop}%
\bibitem [{\citenamefont {Giannozzi}\ \emph {et~al.}(2017)\citenamefont
  {Giannozzi}, \citenamefont {Andreussi}, \citenamefont {Brumme}, \citenamefont
  {Bunau}, \citenamefont {Nardelli}, \citenamefont {Calandra}, \citenamefont
  {Car}, \citenamefont {Cavazzoni}, \citenamefont {Ceresoli}, \citenamefont
  {Cococcioni}, \citenamefont {Colonna}, \citenamefont {Carnimeo},
  \citenamefont {Corso}, \citenamefont {de~Gironcoli}, \citenamefont {Delugas},
  \citenamefont {Jr}, \citenamefont {Ferretti}, \citenamefont {Floris},
  \citenamefont {Fratesi}, \citenamefont {Fugallo}, \citenamefont {Gebauer},
  \citenamefont {Gerstmann}, \citenamefont {Giustino}, \citenamefont {Gorni},
  \citenamefont {Jia}, \citenamefont {Kawamura}, \citenamefont {Ko},
  \citenamefont {Kokalj}, \citenamefont {Küçükbenli}, \citenamefont
  {Lazzeri}, \citenamefont {Marsili}, \citenamefont {Marzari}, \citenamefont
  {Mauri}, \citenamefont {Nguyen}, \citenamefont {Nguyen}, \citenamefont {de-la
  Roza}, \citenamefont {Paulatto}, \citenamefont {Poncé}, \citenamefont
  {Rocca}, \citenamefont {Sabatini}, \citenamefont {Santra}, \citenamefont
  {Schlipf}, \citenamefont {Seitsonen}, \citenamefont {Smogunov}, \citenamefont
  {Timrov}, \citenamefont {Thonhauser}, \citenamefont {Umari}, \citenamefont
  {Vast}, \citenamefont {Wu},\ and\ \citenamefont {Baroni}}]{QE17}%
  \BibitemOpen
  \bibfield  {author} {\bibinfo {author} {\bibfnamefont {P.}~\bibnamefont
  {Giannozzi}}, \bibinfo {author} {\bibfnamefont {O.}~\bibnamefont
  {Andreussi}}, \bibinfo {author} {\bibfnamefont {T.}~\bibnamefont {Brumme}},
  \bibinfo {author} {\bibfnamefont {O.}~\bibnamefont {Bunau}}, \bibinfo
  {author} {\bibfnamefont {M.~B.}\ \bibnamefont {Nardelli}}, \bibinfo {author}
  {\bibfnamefont {M.}~\bibnamefont {Calandra}}, \bibinfo {author}
  {\bibfnamefont {R.}~\bibnamefont {Car}}, \bibinfo {author} {\bibfnamefont
  {C.}~\bibnamefont {Cavazzoni}}, \bibinfo {author} {\bibfnamefont
  {D.}~\bibnamefont {Ceresoli}}, \bibinfo {author} {\bibfnamefont
  {M.}~\bibnamefont {Cococcioni}}, \bibinfo {author} {\bibfnamefont
  {N.}~\bibnamefont {Colonna}}, \bibinfo {author} {\bibfnamefont
  {I.}~\bibnamefont {Carnimeo}}, \bibinfo {author} {\bibfnamefont {A.~D.}\
  \bibnamefont {Corso}}, \bibinfo {author} {\bibfnamefont {S.}~\bibnamefont
  {de~Gironcoli}}, \bibinfo {author} {\bibfnamefont {P.}~\bibnamefont
  {Delugas}}, \bibinfo {author} {\bibfnamefont {R.~A.~D.}\ \bibnamefont {Jr}},
  \bibinfo {author} {\bibfnamefont {A.}~\bibnamefont {Ferretti}}, \bibinfo
  {author} {\bibfnamefont {A.}~\bibnamefont {Floris}}, \bibinfo {author}
  {\bibfnamefont {G.}~\bibnamefont {Fratesi}}, \bibinfo {author} {\bibfnamefont
  {G.}~\bibnamefont {Fugallo}}, \bibinfo {author} {\bibfnamefont
  {R.}~\bibnamefont {Gebauer}}, \bibinfo {author} {\bibfnamefont
  {U.}~\bibnamefont {Gerstmann}}, \bibinfo {author} {\bibfnamefont
  {F.}~\bibnamefont {Giustino}}, \bibinfo {author} {\bibfnamefont
  {T.}~\bibnamefont {Gorni}}, \bibinfo {author} {\bibfnamefont
  {J.}~\bibnamefont {Jia}}, \bibinfo {author} {\bibfnamefont {M.}~\bibnamefont
  {Kawamura}}, \bibinfo {author} {\bibfnamefont {H.-Y.}\ \bibnamefont {Ko}},
  \bibinfo {author} {\bibfnamefont {A.}~\bibnamefont {Kokalj}}, \bibinfo
  {author} {\bibfnamefont {E.}~\bibnamefont {Küçükbenli}}, \bibinfo {author}
  {\bibfnamefont {M.}~\bibnamefont {Lazzeri}}, \bibinfo {author} {\bibfnamefont
  {M.}~\bibnamefont {Marsili}}, \bibinfo {author} {\bibfnamefont
  {N.}~\bibnamefont {Marzari}}, \bibinfo {author} {\bibfnamefont
  {F.}~\bibnamefont {Mauri}}, \bibinfo {author} {\bibfnamefont {N.~L.}\
  \bibnamefont {Nguyen}}, \bibinfo {author} {\bibfnamefont {H.-V.}\
  \bibnamefont {Nguyen}}, \bibinfo {author} {\bibfnamefont {A.~O.}\
  \bibnamefont {de-la Roza}}, \bibinfo {author} {\bibfnamefont
  {L.}~\bibnamefont {Paulatto}}, \bibinfo {author} {\bibfnamefont
  {S.}~\bibnamefont {Poncé}}, \bibinfo {author} {\bibfnamefont
  {D.}~\bibnamefont {Rocca}}, \bibinfo {author} {\bibfnamefont
  {R.}~\bibnamefont {Sabatini}}, \bibinfo {author} {\bibfnamefont
  {B.}~\bibnamefont {Santra}}, \bibinfo {author} {\bibfnamefont
  {M.}~\bibnamefont {Schlipf}}, \bibinfo {author} {\bibfnamefont {A.~P.}\
  \bibnamefont {Seitsonen}}, \bibinfo {author} {\bibfnamefont {A.}~\bibnamefont
  {Smogunov}}, \bibinfo {author} {\bibfnamefont {I.}~\bibnamefont {Timrov}},
  \bibinfo {author} {\bibfnamefont {T.}~\bibnamefont {Thonhauser}}, \bibinfo
  {author} {\bibfnamefont {P.}~\bibnamefont {Umari}}, \bibinfo {author}
  {\bibfnamefont {N.}~\bibnamefont {Vast}}, \bibinfo {author} {\bibfnamefont
  {X.}~\bibnamefont {Wu}}, \ and\ \bibinfo {author} {\bibfnamefont
  {S.}~\bibnamefont {Baroni}},\ }\href
  {http://stacks.iop.org/0953-8984/29/i=46/a=465901} {\bibfield  {journal}
  {\bibinfo  {journal} {Journal of Physics: Condensed Matter}\ }\textbf
  {\bibinfo {volume} {29}},\ \bibinfo {pages} {465901} (\bibinfo {year}
  {2017})}\BibitemShut {NoStop}%
\bibitem [{\citenamefont {Corso}(2014)}]{PSLibrary}%
  \BibitemOpen
  \bibfield  {author} {\bibinfo {author} {\bibfnamefont {A.~D.}\ \bibnamefont
  {Corso}},\ }\href {\doibase https://doi.org/10.1016/j.commatsci.2014.07.043}
  {\bibfield  {journal} {\bibinfo  {journal} {Computational Materials Science}\
  }\textbf {\bibinfo {volume} {95}},\ \bibinfo {pages} {337 } (\bibinfo {year}
  {2014})}\BibitemShut {NoStop}%
\bibitem [{\citenamefont {Monkhorst}\ and\ \citenamefont {Pack}(1976)}]{MP76}%
  \BibitemOpen
  \bibfield  {author} {\bibinfo {author} {\bibfnamefont {H.~J.}\ \bibnamefont
  {Monkhorst}}\ and\ \bibinfo {author} {\bibfnamefont {J.~D.}\ \bibnamefont
  {Pack}},\ }\href {\doibase 10.1103/PhysRevB.13.5188} {\bibfield  {journal}
  {\bibinfo  {journal} {Phys. Rev. B}\ }\textbf {\bibinfo {volume} {13}},\
  \bibinfo {pages} {5188} (\bibinfo {year} {1976})}\BibitemShut {NoStop}%
\bibitem [{\citenamefont {Methfessel}\ and\ \citenamefont
  {Paxton}(1989)}]{MP89}%
  \BibitemOpen
  \bibfield  {author} {\bibinfo {author} {\bibfnamefont {M.}~\bibnamefont
  {Methfessel}}\ and\ \bibinfo {author} {\bibfnamefont {A.~T.}\ \bibnamefont
  {Paxton}},\ }\href {\doibase 10.1103/PhysRevB.40.3616} {\bibfield  {journal}
  {\bibinfo  {journal} {Phys. Rev. B}\ }\textbf {\bibinfo {volume} {40}},\
  \bibinfo {pages} {3616} (\bibinfo {year} {1989})}\BibitemShut {NoStop}%
\bibitem [{\citenamefont {Mostofi}\ \emph {et~al.}(2008)\citenamefont
  {Mostofi}, \citenamefont {Yates}, \citenamefont {Lee}, \citenamefont {Souza},
  \citenamefont {Vanderbilt},\ and\ \citenamefont {Marzari}}]{Mostofi2008685}%
  \BibitemOpen
  \bibfield  {author} {\bibinfo {author} {\bibfnamefont {A.~A.}\ \bibnamefont
  {Mostofi}}, \bibinfo {author} {\bibfnamefont {J.~R.}\ \bibnamefont {Yates}},
  \bibinfo {author} {\bibfnamefont {Y.-S.}\ \bibnamefont {Lee}}, \bibinfo
  {author} {\bibfnamefont {I.}~\bibnamefont {Souza}}, \bibinfo {author}
  {\bibfnamefont {D.}~\bibnamefont {Vanderbilt}}, \ and\ \bibinfo {author}
  {\bibfnamefont {N.}~\bibnamefont {Marzari}},\ }\href {\doibase
  http://dx.doi.org/10.1016/j.cpc.2007.11.016} {\bibfield  {journal} {\bibinfo
  {journal} {Computer Physics Communications}\ }\textbf {\bibinfo {volume}
  {178}},\ \bibinfo {pages} {685 } (\bibinfo {year} {2008})}\BibitemShut
  {NoStop}%
\bibitem [{\citenamefont {Kaltak}(2015)}]{Merzuk_PhD}%
  \BibitemOpen
  \bibfield  {author} {\bibinfo {author} {\bibfnamefont {M.}~\bibnamefont
  {Kaltak}},\ }\href {http://othes.univie.ac.at/38099/} {\bibinfo {type}
  {{PhD}. dissertation}},\ \bibinfo  {school} {Universit{\"a}t Wien} (\bibinfo
  {year} {2015})\BibitemShut {NoStop}%
\bibitem [{\citenamefont {R\"osner}\ \emph {et~al.}(2015)\citenamefont
  {R\"osner}, \citenamefont {\ifmmode \mbox{\c{S}}\else \c{S}\fi{}a\ifmmode
  \mbox{\c{s}}\else \c{s}\fi{}\ifmmode \imath \else \i
  \fi{}o\ifmmode~\breve{g}\else \u{g}\fi{}lu}, \citenamefont {Friedrich},
  \citenamefont {Bl\"ugel},\ and\ \citenamefont {Wehling}}]{Malte_wfce}%
  \BibitemOpen
  \bibfield  {author} {\bibinfo {author} {\bibfnamefont {M.}~\bibnamefont
  {R\"osner}}, \bibinfo {author} {\bibfnamefont {E.}~\bibnamefont {\ifmmode
  \mbox{\c{S}}\else \c{S}\fi{}a\ifmmode \mbox{\c{s}}\else \c{s}\fi{}\ifmmode
  \imath \else \i \fi{}o\ifmmode~\breve{g}\else \u{g}\fi{}lu}}, \bibinfo
  {author} {\bibfnamefont {C.}~\bibnamefont {Friedrich}}, \bibinfo {author}
  {\bibfnamefont {S.}~\bibnamefont {Bl\"ugel}}, \ and\ \bibinfo {author}
  {\bibfnamefont {T.~O.}\ \bibnamefont {Wehling}},\ }\href {\doibase
  10.1103/PhysRevB.92.085102} {\bibfield  {journal} {\bibinfo  {journal} {Phys.
  Rev. B}\ }\textbf {\bibinfo {volume} {92}},\ \bibinfo {pages} {085102}
  (\bibinfo {year} {2015})}\BibitemShut {NoStop}%
\bibitem [{\citenamefont {Anisimov}\ \emph {et~al.}(1997)\citenamefont
  {Anisimov}, \citenamefont {Poteryaev}, \citenamefont {Korotin}, \citenamefont
  {Anokhin},\ and\ \citenamefont {Kotliar}}]{anisimov_first-principles_1997}%
  \BibitemOpen
  \bibfield  {author} {\bibinfo {author} {\bibfnamefont {V.~I.}\ \bibnamefont
  {Anisimov}}, \bibinfo {author} {\bibfnamefont {A.~I.}\ \bibnamefont
  {Poteryaev}}, \bibinfo {author} {\bibfnamefont {M.~A.}\ \bibnamefont
  {Korotin}}, \bibinfo {author} {\bibfnamefont {A.~O.}\ \bibnamefont
  {Anokhin}}, \ and\ \bibinfo {author} {\bibfnamefont {G.}~\bibnamefont
  {Kotliar}},\ }\href {\doibase 10.1088/0953-8984/9/35/010} {\bibfield
  {journal} {\bibinfo  {journal} {J. Phys.: Condens. Matter}\ }\textbf
  {\bibinfo {volume} {9}},\ \bibinfo {pages} {7359} (\bibinfo {year}
  {1997})}\BibitemShut {NoStop}%
\bibitem [{\citenamefont {Lichtenstein}\ and\ \citenamefont
  {Katsnelson}(1998)}]{lichtenstein__1998}%
  \BibitemOpen
  \bibfield  {author} {\bibinfo {author} {\bibfnamefont {A.~I.}\ \bibnamefont
  {Lichtenstein}}\ and\ \bibinfo {author} {\bibfnamefont {M.~I.}\ \bibnamefont
  {Katsnelson}},\ }\href {\doibase 10.1103/PhysRevB.57.6884} {\bibfield
  {journal} {\bibinfo  {journal} {Phys. Rev. B}\ }\textbf {\bibinfo {volume}
  {57}},\ \bibinfo {pages} {6884} (\bibinfo {year} {1998})}\BibitemShut
  {NoStop}%
\bibitem [{\citenamefont {Georges}\ \emph {et~al.}(1996)\citenamefont
  {Georges}, \citenamefont {Kotliar}, \citenamefont {Krauth},\ and\
  \citenamefont {Rozenberg}}]{georges_dynamical_1996}%
  \BibitemOpen
  \bibfield  {author} {\bibinfo {author} {\bibfnamefont {A.}~\bibnamefont
  {Georges}}, \bibinfo {author} {\bibfnamefont {G.}~\bibnamefont {Kotliar}},
  \bibinfo {author} {\bibfnamefont {W.}~\bibnamefont {Krauth}}, \ and\ \bibinfo
  {author} {\bibfnamefont {M.~J.}\ \bibnamefont {Rozenberg}},\ }\href {\doibase
  10.1103/RevModPhys.68.13} {\bibfield  {journal} {\bibinfo  {journal} {Rev.
  Mod. Phys.}\ }\textbf {\bibinfo {volume} {68}},\ \bibinfo {pages} {13}
  (\bibinfo {year} {1996})}\BibitemShut {NoStop}%
\bibitem [{\citenamefont {Gull}\ \emph {et~al.}(2011)\citenamefont {Gull},
  \citenamefont {Millis}, \citenamefont {Lichtenstein}, \citenamefont
  {Rubtsov}, \citenamefont {Troyer},\ and\ \citenamefont
  {Werner}}]{EGull_ctqmc_rmp}%
  \BibitemOpen
  \bibfield  {author} {\bibinfo {author} {\bibfnamefont {E.}~\bibnamefont
  {Gull}}, \bibinfo {author} {\bibfnamefont {A.~J.}\ \bibnamefont {Millis}},
  \bibinfo {author} {\bibfnamefont {A.~I.}\ \bibnamefont {Lichtenstein}},
  \bibinfo {author} {\bibfnamefont {A.~N.}\ \bibnamefont {Rubtsov}}, \bibinfo
  {author} {\bibfnamefont {M.}~\bibnamefont {Troyer}}, \ and\ \bibinfo {author}
  {\bibfnamefont {P.}~\bibnamefont {Werner}},\ }\href {\doibase
  10.1103/RevModPhys.83.349} {\bibfield  {journal} {\bibinfo  {journal} {Rev.
  Mod. Phys.}\ }\textbf {\bibinfo {volume} {83}},\ \bibinfo {pages} {349}
  (\bibinfo {year} {2011})}\BibitemShut {NoStop}%
\bibitem [{\citenamefont {Parragh}\ \emph
  {et~al.}(2012{\natexlab{b}})\citenamefont {Parragh}, \citenamefont {Toschi},
  \citenamefont {Held},\ and\ \citenamefont
  {Sangiovanni}}]{Sagiovani_prb2012_conservedquant}%
  \BibitemOpen
  \bibfield  {author} {\bibinfo {author} {\bibfnamefont {N.}~\bibnamefont
  {Parragh}}, \bibinfo {author} {\bibfnamefont {A.}~\bibnamefont {Toschi}},
  \bibinfo {author} {\bibfnamefont {K.}~\bibnamefont {Held}}, \ and\ \bibinfo
  {author} {\bibfnamefont {G.}~\bibnamefont {Sangiovanni}},\ }\href {\doibase
  10.1103/PhysRevB.86.155158} {\bibfield  {journal} {\bibinfo  {journal} {Phys.
  Rev. B}\ }\textbf {\bibinfo {volume} {86}},\ \bibinfo {pages} {155158}
  (\bibinfo {year} {2012}{\natexlab{b}})}\BibitemShut {NoStop}%
\bibitem [{\citenamefont {Prokof'ev}\ \emph {et~al.}(1998)\citenamefont
  {Prokof'ev}, \citenamefont {Svistunov},\ and\ \citenamefont
  {Tupitsyn}}]{worm1998}%
  \BibitemOpen
  \bibfield  {author} {\bibinfo {author} {\bibfnamefont {N.~V.}\ \bibnamefont
  {Prokof'ev}}, \bibinfo {author} {\bibfnamefont {B.~V.}\ \bibnamefont
  {Svistunov}}, \ and\ \bibinfo {author} {\bibfnamefont {I.~S.}\ \bibnamefont
  {Tupitsyn}},\ }\href {\doibase 10.1134/1.558661} {\bibfield  {journal}
  {\bibinfo  {journal} {Journal of Experimental and Theoretical Physics}\
  }\textbf {\bibinfo {volume} {87}},\ \bibinfo {pages} {310} (\bibinfo {year}
  {1998})}\BibitemShut {NoStop}%
\bibitem [{\citenamefont {Gunacker}\ \emph {et~al.}(2015)\citenamefont
  {Gunacker}, \citenamefont {Wallerberger}, \citenamefont {Gull}, \citenamefont
  {Hausoel}, \citenamefont {Sangiovanni},\ and\ \citenamefont
  {Held}}]{sangiovani_prb2015_worm}%
  \BibitemOpen
  \bibfield  {author} {\bibinfo {author} {\bibfnamefont {P.}~\bibnamefont
  {Gunacker}}, \bibinfo {author} {\bibfnamefont {M.}~\bibnamefont
  {Wallerberger}}, \bibinfo {author} {\bibfnamefont {E.}~\bibnamefont {Gull}},
  \bibinfo {author} {\bibfnamefont {A.}~\bibnamefont {Hausoel}}, \bibinfo
  {author} {\bibfnamefont {G.}~\bibnamefont {Sangiovanni}}, \ and\ \bibinfo
  {author} {\bibfnamefont {K.}~\bibnamefont {Held}},\ }\href {\doibase
  10.1103/PhysRevB.92.155102} {\bibfield  {journal} {\bibinfo  {journal} {Phys.
  Rev. B}\ }\textbf {\bibinfo {volume} {92}},\ \bibinfo {pages} {155102}
  (\bibinfo {year} {2015})}\BibitemShut {NoStop}%
\bibitem [{\citenamefont {Gunacker}\ \emph {et~al.}(2016)\citenamefont
  {Gunacker}, \citenamefont {Wallerberger}, \citenamefont {Ribic},
  \citenamefont {Hausoel}, \citenamefont {Sangiovanni},\ and\ \citenamefont
  {Held}}]{sangiovani_prb2016_worm}%
  \BibitemOpen
  \bibfield  {author} {\bibinfo {author} {\bibfnamefont {P.}~\bibnamefont
  {Gunacker}}, \bibinfo {author} {\bibfnamefont {M.}~\bibnamefont
  {Wallerberger}}, \bibinfo {author} {\bibfnamefont {T.}~\bibnamefont {Ribic}},
  \bibinfo {author} {\bibfnamefont {A.}~\bibnamefont {Hausoel}}, \bibinfo
  {author} {\bibfnamefont {G.}~\bibnamefont {Sangiovanni}}, \ and\ \bibinfo
  {author} {\bibfnamefont {K.}~\bibnamefont {Held}},\ }\href {\doibase
  10.1103/PhysRevB.94.125153} {\bibfield  {journal} {\bibinfo  {journal} {Phys.
  Rev. B}\ }\textbf {\bibinfo {volume} {94}},\ \bibinfo {pages} {125153}
  (\bibinfo {year} {2016})}\BibitemShut {NoStop}%
\bibitem [{\citenamefont {{Sch{\"u}ler}}\ \emph {et~al.}(2018)\citenamefont
  {{Sch{\"u}ler}}, \citenamefont {{Peil}}, \citenamefont {{Kraberger}},
  \citenamefont {{Pordzik}}, \citenamefont {{Marsman}}, \citenamefont
  {{Kresse}}, \citenamefont {{Wehling}},\ and\ \citenamefont
  {{Aichhorn}}}]{Schulerprojector2018}%
  \BibitemOpen
  \bibfield  {author} {\bibinfo {author} {\bibfnamefont {M.}~\bibnamefont
  {{Sch{\"u}ler}}}, \bibinfo {author} {\bibfnamefont {O.~E.}\ \bibnamefont
  {{Peil}}}, \bibinfo {author} {\bibfnamefont {G.~J.}\ \bibnamefont
  {{Kraberger}}}, \bibinfo {author} {\bibfnamefont {R.}~\bibnamefont
  {{Pordzik}}}, \bibinfo {author} {\bibfnamefont {M.}~\bibnamefont
  {{Marsman}}}, \bibinfo {author} {\bibfnamefont {G.}~\bibnamefont {{Kresse}}},
  \bibinfo {author} {\bibfnamefont {T.~O.}\ \bibnamefont {{Wehling}}}, \ and\
  \bibinfo {author} {\bibfnamefont {M.}~\bibnamefont {{Aichhorn}}},\
  }\href@noop {} {\bibfield  {journal} {\bibinfo  {journal} {ArXiv e-prints}\ }
  (\bibinfo {year} {2018})},\ \Eprint {http://arxiv.org/abs/1804.02055}
  {arXiv:1804.02055 [cond-mat.str-el]} \BibitemShut {NoStop}%
\end{thebibliography}%
\end{document}